# ONE-DIMENSIONAL QUANTILE-STRATIFIED SAMPLING AND ITS APPLICATION IN STATISTICAL SIMULATIONS

BEN O'NEILL,[*] *ACIL Allen*[**]

**Abstract**

In this paper we examine quantile-stratified samples from a known univariate probability distribution, with stratification occurring over a partition of the quantile regions in the distribution. We examine some general properties of this sampling method and we contrast it with standard IID sampling to highlight its similarities and differences. We examine the applications of this sampling method to various statistical simulations including importance sampling. We conduct simulation analysis to compare the performance of standard importance sampling against the quantile-stratified importance sampling to see how they each perform on a range of functions.

## 1. Introduction

Pseudo-random sampling from known probability distributions is used widely in mathematical and statistical applications. Random sampling of this kind is used in simulation analysis and for a range of statistical simulation methods, including importance sampling and other Monte-Carlo methods. Many procedures of this kind use IID samples from the stipulated probability distribution, but the convergence of the estimation procedures they are applied to often make use of ergodic theorems that do not require the underlying sample values to be independent.[1]

One aspect of the standard IID sample that is relevant to some estimation problems is the degree to which the empirical quantile function of the generated sample approximates the true quantile function of the underlying sampling distribution. Well-known results for order statistics in IID samples can be invoked to understand this correspondence in detail, but generally speaking, there is a reasonable amount of random variation in the sample quantiles in an IID sample and this means that the empirical quantiles may approximate the true quantiles a bit less closely than we would ideally like them to when constructing estimators. More broadly, it is often useful for the random sampling design to be "space filling" in the sense that the generated random sample will tend to cover most "areas in the support" of the probability distribution at issue. Various kinds of "space-filling designs" have been used in design methods for random and deterministic sampling methods to achieve this goal (see e.g., Santner, Williams and Notz 2003, Ch 5; Draguljic, Santner and Dean 2012). The most widely used space-filling design is the Latin hypercube method (see e.g., Street and Street 1987; Tang 1993; Cox and Reid 2000, p. 181-182; Hinkelmann 2011, 401-419; Iman 2013) but it also includes distance-based designs such as maximin and minimax designs (Johnson, Moore and Ylvisaker 1990) and more recent "fast, flexible filling" designs for irregular or constrained spaces (Lekivetz and Jones 2014).

[*] E-mail address: ben.oneill@hotmail.com.
[**] Level 6, 54 Marcus Clarke Street, Canberra ACT 2601, Australia.

[1] Broader Monte-Carlo Markov Chain (MCMC) methods typically do not involve independent simulations and instead simulate from Markov chains with autocorrelation between the values in the chain.

Standard IID random sampling tends not to have particularly good "space-filling" properties and tends not to estimate distribution quantiles very well. One reasonable method to improve these aspects of the random sampling method is to use "stratification" over a set of strata that partition the support of the sampling distribution, thereby "filling the space" by judicious use of strata. While different strata could —in principle— be used for this task, the most obvious way to do this is to break up the support into a set of equiprobable regions based on the quantiles of the sampling distribution. Santner, Williams and Notz (2003) have noted that stratification can be used to obtain a space-filling design, observing that, "[t]o improve the chances that inputs are spread "evenly" over the experimental region, … we want a design consisting of $n$ points, we would divide the experimental region into $n$ strata, spread evenly throughout the experimental region, and randomly select a single point from each" (p. 114).

Stratified sampling methods of this kind have been used in sampling problems in the fields of hydrology, meteorology and ecology (see e.g., Claggett et al. 2010, Wallenius et al. 2011, Noble et al. 2012, Padilla et al. 2014, Ding and Lee 2014, Hu et al. 2016). The method at issue is often simply called "stratified sampling", but we will call the method "quantile-stratified sampling" to be specific about the method of stratification. This is a "space-filling" method of sampling which generates pseudo-random samples that maintain the desired (marginal) sampling distribution. The method can be thought of as a one-dimensional case of randomised Latin hypercube sampling, where each probability region use for the stratification is just an equal-sized line segment of the unit interval. While quantile-stratified sampling is equivalent to one-dimensional Latin hypercube sampling, it can be extended to higher dimensional cases in ways that depart from the Latin hypercube method, making it a base case worthy of special examination in its own right.

In the present paper we will describe quantile-stratified sampling for one-dimensional sampling and analyse its properties and its potential applications. We describe the method and compare it to IID sampling using an analogy to simple random sampling with and without replacement, and we examine the effect of stratification relative to IID sampling. Although the use of stratification for random sampling is not new, our characterisation of the method is novel and provides a useful analogy to sampling with and without replacement in discrete settings. We also examine a layered variant of the sampling method that is novel, providing an intermediate case of interest. We will derive properties of quantile-stratified sampling and connect these properties to broader results for Latin hypercube sampling, including examining a variation of importance sampling using a quantile-stratified sample from the candidate distribution. Results relating to estimation properties of this kind are general asymptotic and they depend heavily on the functions at issue, so we undertake simulation analysis over a range of standard problems to test the performance of quantile-stratified sampling and IID sampling in this context. To do this we generate simulation data over combinations of different proposal/target distributions and we undertake regression analysis on the resulting simulation data. We find evidence that QS sampling reduces estimation error relative to IID sampling, and the improvement in the resulting estimator increases for higher sample sizes. Finally, we provide discussion on how one-dimensional quantile-stratified sampling could fruitfully be generalised to distributions in higher dimensions, in ways that do not follow the Latin hypercube approach, but which follow certain natural transformation properties of multivariate distributions.

To avoid confusion, it is worth also stressing what we are **not** doing. The present paper is not about using stratified sampling to estimate an unknown distribution or quantiles, or performing any kind of inference relating to the sampling distribution. There is already a large statistical literature on inference and estimation problems for unknown distributions using stratified samples and this is irrelevant to our purposes. Here we will assume that the stipulated sampling distribution is known, with a known and computable quantile function, and that it is consequently possible to generate values from the distribution (or any conditional part of that distribution) through the standard inverse transformation algorithm. Our focus here will be on determining the behaviours of IID and quantile-stratified samples from a known univariate distribution. Our overarching purpose is to investigate the use of this method of “space-filling” univariate random sampling and compare it to IID sampling. In later research the method will be generalised to provide “space-filling” multivariate random sampling methods.

## 2. Quantile-stratified sampling versus IID sampling

Suppose we wish to generate pseudo-random values from a univariate distribution with known density function $f$, distribution function $F$ and quantile function $Q$. There are a few different ways to define the sampling method of interest here, but we will describe it in a non-standard way that elucidates its connection with simple-random-sampling of values from a finite population. To sample $n$ values from the distribution we first break the support of the distribution up into equiprobable quantile-blocks using the quantile function. We can generate an **IID sample** or a **quantile-stratified (QS) sample** by generating simple random samples with or without replacement over the indices for those quantile-blocks and then sampling over the conditional distribution over each of the quantile-blocks:

$$\text{IID:} \qquad S_1, \dots, S_n \sim \text{SRSWR}\{1, \dots, n\} \qquad U_i \sim \text{U}\left(\frac{S_i - 1}{n}, \frac{S_i}{n}\right) \qquad X_i \equiv Q(U_i),$$

$$\text{QS:} \qquad S_1^*, \dots, S_n^* \sim \text{SRSWOR}\{1, \dots, n\} \qquad U_i^* \sim \text{U}\left(\frac{S_i^* - 1}{n}, \frac{S_i^*}{n}\right) \qquad X_i^* \equiv Q(U_i^*).$$

(Note that this is a non-standard way of presenting IID sampling, but it is simple to verify that this method will yield independent values from the sampling distribution. We present it this way to show the similarity to QS sampling.) Throughout the remainder of the paper, we will refer to these processes using the following simple shorthand:

$$X_1, \dots, X_n \sim \text{IID } f,$$

$$X_1^*, \dots, X_n^* \sim \text{QS } f.$$

It is simple to establish that both methods yield a marginal distribution equal to the desired sampling distribution, with the former method having independent values and the latter method having dependent values. To see that both methods give the desired sampling distribution, suppose we let $Q(0) = w_0 < w_1 < \cdots < w_n = Q(1)$ denote the relevant quantiles of the distribution (which are the boundaries of the quantile-blocks), given by:

$$w_s \equiv Q\left(\frac{s}{n}\right).$$

If the sampling distribution is continuous then conditional on the selection of the quantile-block $s$, we sample over the conditional distribution of the sampling distribution over the interval $[w_{s-1}, w_s]$ and the resulting conditional density function, cumulative distribution function and quantile function are given respectively by:

$$f(x|s) = n \cdot f(x) \cdot \mathbb{I}(w_{s-1} < x \leq w_s),$$

$$F(x|s) = \mathbb{I}(x > w_s) + n \cdot \mathbb{I}(w_{s-1} < x \leq w_s)\left(F(x) - \frac{s-1}{n}\right),$$

$$Q(p|s) = Q\left(\frac{s+p-1}{n}\right).$$

Using the law of total probability we have:

$$\begin{aligned}
\mathbb{P}(X_i^* \leq x) &= \sum_{s=1}^{n} \mathbb{P}(X_i^* \leq x | S_i^* = s) \cdot \mathbb{P}(S_i^* = s) \\
&= \frac{1}{n}\sum_{s=1}^{n} \mathbb{P}(X_i^* \leq x | S_i^* = s) \\
&= \frac{1}{n}\sum_{s=1}^{n} \int_{-\infty}^{x} f(r|s)\, dr \\
&= \sum_{s=1}^{n} \int_{-\infty}^{x} f(r) \cdot \mathbb{I}(w_{s-1} < r \leq w_s)\, dr \\
&= \int_{-\infty}^{x} f(r) \left( \sum_{i=1}^{n} \mathbb{I}(w_{s-1} < r \leq w_s) \right) dr \\
&= \int_{-\infty}^{x} f(r) \cdot \mathbb{I}(Q(0) < r \leq Q(1))\, dr \\
&= \int_{-\infty}^{x} f(r)\, dr \\
&= F(x),
\end{aligned}$$

and the corresponding demonstration for the distribution of $X_i$ follows analogously. The case for a non-continuous distribution is a bit more complicated and may involve some splitting of outcomes occurring with non-zero probability at the boundary point of the quantile-block, but the sampling method still works, and the desired sampling distribution still holds.

**REMARK:** The quantile-stratified sampling method and the framing of the sampling methods works even if the distribution function is non-continuous (e.g., has jump point). This is because inverse transformation sampling still works for non-continuous distributions. If there is a jump point at the boundary of quantile-blocks, this may occur with non-zero probability in both of the quantile-blocks that share that boundary. □

As can be seen from above, the difference between IID sampling and QS sampling is analogous to the difference between SRSWR and SRSWOR. In the former case, each sample value comes from a random quantile-block and the occurrence of a previous observation in a quantile-block does not alter this.[2] Consequently, the values in the former case are independent and so they are an IID sample from the specified sampling distribution. In the latter case, once a value is obtained from a quantile-block, there is no "replacement" of that quantile-block and subsequent values must come from other quantile-blocks. This induces negative correlation between the QS uniform variables in the latter method, as shown in Theorem 1 below. As the sample size for become large, this negative correlation vanishes and the two methods converge. (Note that the IID sample values have the same mean and variance as this, but they are uncorrelated.)

**THEOREM 1:** The QS uniform random variables $U_1^*, \dots, U_n^*$ have moments:

$$\mathbb{E}(U_i^*) = \frac{1}{2}$$

$$\mathbb{V}(U_i^*) = \frac{1}{12}$$

$$\mathbb{C}(U_i^*, U_j^*) = -\frac{n+1}{12n^2} \qquad i \neq j.$$

$$\mathbb{Corr}(U_i^*, U_j^*) = -\frac{n+1}{n^2} \qquad i \neq j.$$

Generally speaking, the negative correlation in the QS uniform random variables flow through to the QS sample from the sampling distribution, but since this is a nonlinear transformation, the resulting correlation is complicated. Although the variables generated by each process have the same marginal distribution (by construction), the QS sample forces a single sample value into each of the quantile-blocks, whereas the IID sample has a varying number of values in the quantile-blocks.[3] By forcing the values into these blocks, this means that values in the QS sample typically adheres more closely to the true quantile function than for the IID sample. This is exhibited in Figure 1 below, showing QQ plots for random samples from a standard normal distribution using multiple draws from each type of sampling process.

[2] Our construction of the IID sample by the method shown is not intended to serve as a recommendation for how to generate this sample in practice (since it can be generated far more simply without the construction and sampling from the quantile-blocks. Instead, this characterisation of IID sampling serves to make the contrast with quantile-stratified sampling clearer and show their analogy to SRSWR and SRSWOR.

[3] The vector of counts over the quantile-blocks follows a multinomial distribution with uniform probabilities.

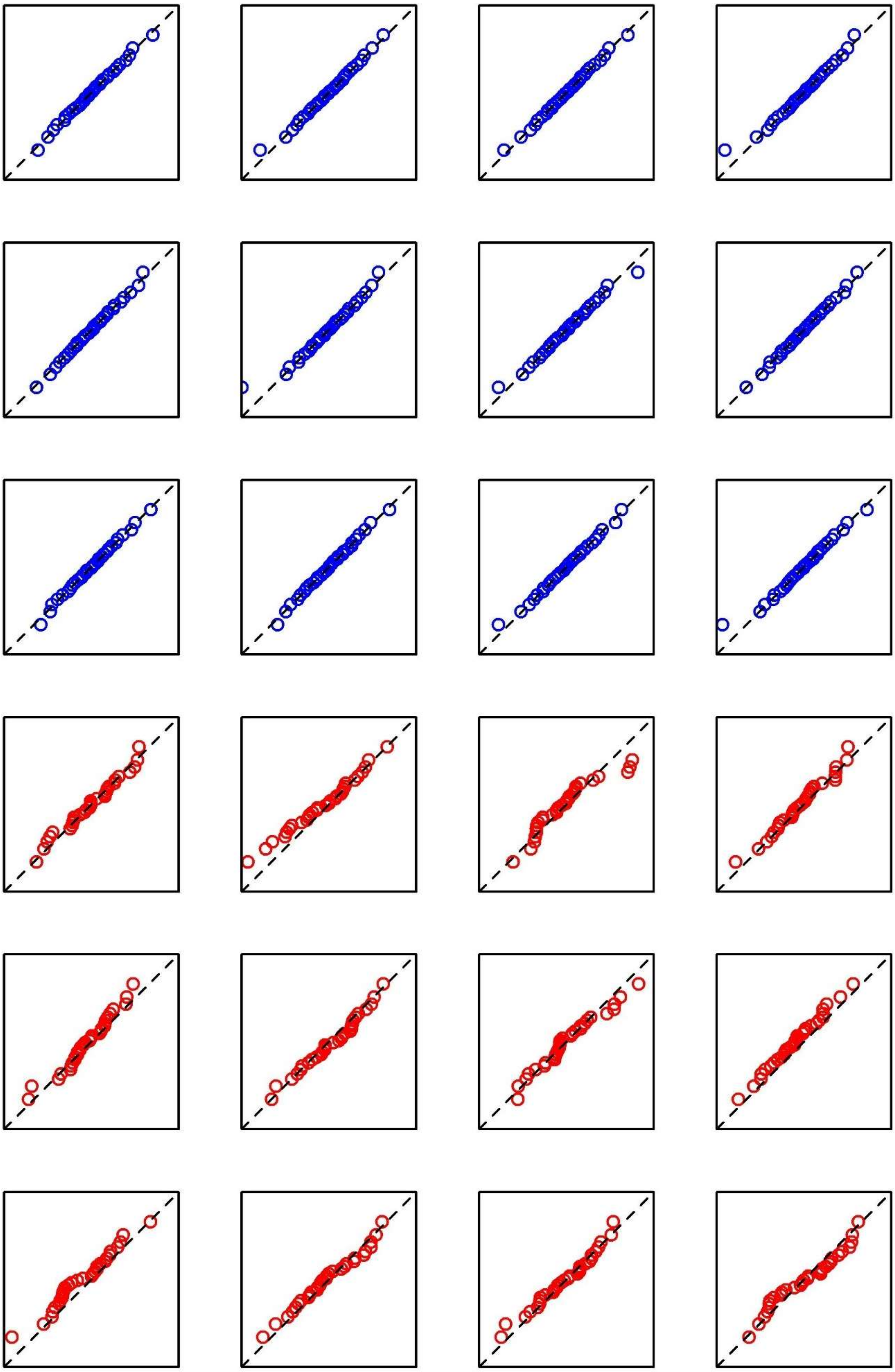

**FIGURE 1:** QQ plots for samples of $n = 30$ data points from standard normal distribution (QS sampling in blue — IID sampling in red)

Whilst not shown in the figure, if the quantile function for the underlying sampling distribution is continuous then the quantile-stratified sample also tends to have a "smoother" empirical quantile function than for an IID sample, owing to its greater adherence to the underlying smoothness of the true quantile function.

Because it approximates the true quantile function of the underlying distribution more closely, there are certain purposes for which a quantile-stratified sample may be a useful substitute to a standard IID sample. In this paper we will use quantile-stratified sampling for the purpose of importance sampling, to estimate the expected value of a function of a random variable (i.e., an integral taken with respect to a probability measure). However, just as there are contexts where a quantile-stratified sample is well-suited, it is also important to recognise that there are also some types of problem for which a quantile-stratified sample is *ill-suited*. In particular, a quantile-stratified sample systematically understates the true variance of the interval between order statistics in an IID random sample (since it forces one value into each quantile-block), so it should not be used as a substitute to an IID random sample for any purposes where the variation of the distance between order statistics must be faithful to IID random sampling. There may also be other areas where the method is ill-suited, and in general, practitioners should analyse the statistical properties of any relevant estimation method that uses quantile-stratified sampling (as we will for its use in importance sampling) before using it as a substitute for IID sampling.

Quantile-stratified sampling has been examined in previous literature (see e.g., Hu et al. 2016), though it is usually just called "stratified sampling" without specification that the stratification method uses the quantile-blocks of the sampling distribution. The method also occurs as the one-dimensional case of Latin hypercube sampling (Loh 2008, Iman 2013), again with the stipulation that the sampling regions are determined by the quantile-blocks. (Thus, another reasonable name for the method is "Latin line sampling".)

## 3. Layered quantile-stratified sampling (an intermediate case)

As has been discussed, the distinction between IID sampling and QS sampling is analogous to the distinction between SRSWR and SRSWOR. The latter method has certain advantages in representing the sampling distribution, but it may also be ill-suited to some situations in some sampling contexts. Depending on the particular application at issue, it might be reasonable to consider the varying properties of IID sampling and QS sampling to constitute a “trade-off”, with greater adherence to the sampling distribution being a desirable property but the negative correlation between values (or some other property of the latter) considered as an undesirable property. If the properties of the two methods are considered to trade-off against one another in terms of their suitability to some application, then it may be reasonable to seek out a “middle ground” between IID sampling and QS sampling.

It is possible to bridge the gap between these two methods of sampling by combining multiple independent subsamples from SRSWOR into a single sample, thereby weakening the negative correlation between the values in the overall sample. In the context of QS sampling, we will refer to this technique as “layering” since each subsample will involve splitting the support of the sampling distribution into a set of quantile blocks, and these partitions of the support can be viewed as a set of “layers” over the support, with one subsample generated for each layer.

We will now describe the technique of “layered quantile sampling” and examine its properties relative to IID sampling and “pure” QS sampling (i.e., QS sampling without layering). We can generate a **layered quantile-stratified (LQS) sample** by generating $K$ subsamples using QS sampling with respective layer sizes $\mathbf{n} = (n_1, \dots, n_K)$ given by:

$$S_{1,1}^{**}, \dots, S_{1,n_1}^{**} \sim \text{SRSWOR}\{1, \dots, n_1\} \qquad U_{k,i}^{**} \sim \text{U}\left(\frac{S_{1,i}^{**} - 1}{n_1}, \frac{S_{1,i}^{**}}{n_1}\right) \qquad X_{1,i}^{**} \equiv Q(U_{1,i}^{*}),$$

$$S_{2,1}^{**}, \dots, S_{2,n_2}^{**} \sim \text{SRSWOR}\{1, \dots, n_2\} \qquad U_{2,i}^{**} \sim \text{U}\left(\frac{S_{2,i}^{**} - 1}{n_2}, \frac{S_{2,i}^{**}}{n_2}\right) \qquad X_{2,i}^{**} \equiv Q(U_{2,i}^{*}),$$

$$\vdots \qquad\qquad \vdots \qquad\qquad \vdots$$

$$S_{K,1}^{**}, \dots, S_{K,n_K}^{**} \sim \text{SRSWOR}\{1, \dots, n_K\} \qquad U_{K,i}^{**} \sim \text{U}\left(\frac{S_{K,i}^{**} - 1}{n_K}, \frac{S_{K,i}^{**}}{n_K}\right) \qquad X_{K,i}^{**} \equiv Q(U_{K,i}^{*}).$$

We then combine and randomly permute these subsamples into a single overall sample to give the LQS sample with $n = n_1 + \cdots + n_K$ sample values:

$$S_1^{**}, \dots, S_n^{**} \sim \text{SRSWOR}\left\{\begin{matrix} (1,1), \dots, (1, n_1), \\ (2,1), \dots, (2, n_2), \\ \vdots \\ (K, 1), \dots, (K, n_K), \end{matrix}\right\} \qquad U_i^{**} \equiv U_{S_i^{**}}^{**} \qquad X_i^{**} \equiv X_{S_i^{**}}^{**}.$$

Throughout the remainder of the paper, we will refer to this process with the shorthand:

$$X_1^{**}, \dots, X_n^{**} \sim \text{LQS}_{\mathbf{n}}\, f.$$

**REMARK:** As with QS sampling, the layered quantile-stratified sampling method also works for non-continuous distributions. To avoid redundancy and reduction of the problem to simpler terms, we stipulate that each value $n_k \geq 1$ so that all the subsamples are non-empty. The value $n$ is considered to be a prespecified size for the LQS sample so the vector $\mathbf{n}$ is chosen subject to these restrictions (i.e., it must be a vector of positive integers that sum to $n$). □

**REMARK:** The LQS sampling method shown above encompasses both IID sampling and QS sampling as special cases (with many cases in between). In the special case where we take $K = n$ (which then gives $n_1 = \cdots = n_K = 1$) each independent subsample has a single value and so the LQS sample is an IID sample. In the special case where we take $K = 1$ (which then gives $n_1 = n$) there is only one subsample, so the LQS sample is just a QS sample. The cases where $1 < K < n$ are the intermediate cases where LQS sampling bridges the gap between IID sampling and QS sampling. □

The difference between LQS sampling and the edge cases of IID sampling and QS sampling is that it combined independent subsamples of generated through SRSWOR. In the non-reductive cases where $1 < K < n$ this means that it is neither fully SRSWR nor SRSWOR, but a mixture of the two sampling types. The presence of some SRSWOR still induces negative correlation between the layered quantile-stratified uniform random variables, but that correlation is now reduced owing to the independence of subsamples, leading to the generalised correlation result shown in Theorem 2 below. It is easily seen that this is a generalisation of Theorem 1, which corresponds to the special case where $K = 1$. As the sample size becomes large in each layer of the sampling method, this negative correlation vanishes and the methods converge.

**THEOREM 2:** The LQS uniform random variables $U_1^{**}, \ldots, U_n^{**}$ have moments:

$$\mathbb{E}(U_i^{**}) = \frac{1}{2}$$

$$\mathbb{V}(U_i^{**}) = \frac{1}{12}$$

$$\mathbb{C}(U_i^{**}, U_j^{**}) = -\frac{n - \sum_{k=1}^{K} 1/n_k}{12n(n-1)} \qquad i \neq j.$$

$$\mathbb{Corr}(U_i^{**}, U_j^{**}) = -\frac{n - \sum_{k=1}^{K} 1/n_k}{n(n-1)} \qquad i \neq j.$$

**COROLLARY:** In the special case where we have $K = n$ layers of sizes $n_1 = \cdots = n_K = 1$ we have $\mathbb{Corr}(U_i^{**}, U_j^{**}) = 0$ for all $i \neq j$ which is the result of an IID sample.

We can relate the correlation results in Theorems 1-2 by the fact that (for $i \neq j$):

$$\begin{aligned}
\mathbb{Corr}(U_i^{**}, U_j^{**}) &= -\frac{n - \sum_{k=1}^{K} 1/n_k}{n(n-1)} \\
&= -\frac{n+1}{n^2} \cdot \frac{n^2 - \sum_{k=1}^{K} n/n_k}{(n+1)(n-1)} \\
&= -\frac{n+1}{n^2} \cdot \frac{n^2 - \sum_{k=1}^{K} n/n_k}{n^2 - 1} \\
&= \mathbb{Corr}(U_i^{*}, U_j^{*}) \cdot \frac{n^2 - \sum_{k=1}^{K} n/n_k}{n^2 - 1} \\
&= \mathbb{Corr}(U_i^{*}, U_j^{*}) \cdot \text{ADJ}(\mathbf{n}),
\end{aligned}$$

using the adjustment term:

$$\text{ADJ}(\mathbf{n}) \equiv \frac{n^2 - \sum_{k=1}^{K} n/n_k}{n^2 - 1}.$$

We can see from the adjustment term that it is heavily affected by the sum of reciprocals of the layer sizes. Having layers with small layer sizes in the LQS sample tends to reduce the negative correlation that is present in pure QS sampling and bring it closer to the independence in IID sampling (with the extreme case where all layers have unit size yielding IID sampling).

Again, the negative correlation in the layered quantile-stratified uniform random variables will generally flow through to the layered quantile-stratified sample from the sampling distribution, but since this is based on a nonlinear transformation the resulting correlation is complicated. The layered quantile-stratified sample forces values in the subsamples into each quantile-block, but if these subsamples have different sizes then the resulting quantile-blocks will not generally correspond and so there is usually some random variation in the number of sample values that fall within any given region. This means that the layered quantile-stratified sample generally adheres more closely to the true quantile function than values in the IID sample but less closely than values in a "pure" quantile-stratified sample. We exhibited this comparison for all three sampling methods in Figure 2 below, showing QQ plots for some randomly generated samples from a standard normal distribution using multiple draws from each sampling method. In this case we have used $K = 3$ layers with layer sizes $\mathbf{n} = (18, 9, 3)$ for the LQS sample, which gives an intrasample correlation of $\mathbb{Corr}(U_i^{**}, U_j^{**}) = -0.03390805$.

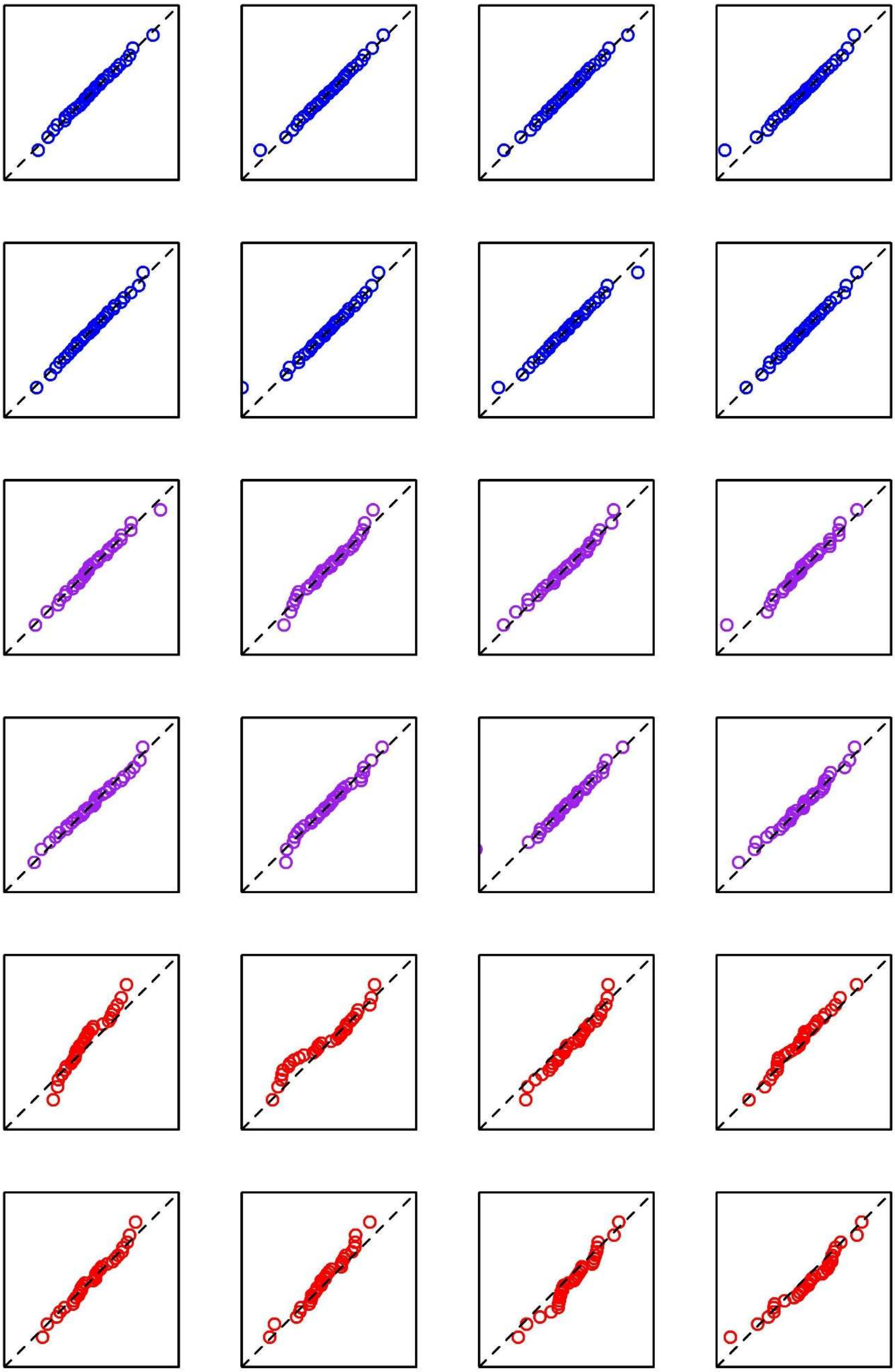

**FIGURE 2:** QQ plots for samples of $n = 30$ data points from standard normal distribution (QS sampling in blue — three-layer LQS sampling in purple — IID sampling in red)

## 4. Comparison of order statistics from the sampling methods

In Figures 1-2 above we established informally that empirical quantiles from quantile-stratified sample tend to adhere better to the true quantiles of the sampling distribution than for the IID sample. There are various ways that this result can be formalised, which we will now explore. (For this part we will consider pure QS sampling, not LQS sampling.) We begin by looking at the degree to which the order statistics from both types of sampling methods tend to adhere to the true quantiles of the distribution. The order statistics serve as empirical quantiles in the sample and thereby function as estimators of the true quantiles of the sampling distribution. We will consider two types of quantile probabilities for this purpose, given by the expected values of the order statistics of the respective samples:

$$p_k \equiv \mathbb{E}(U_{(k)}) = \frac{k}{n+1},$$

$$p_k^* \equiv \mathbb{E}(U_{(k)}^*) = \frac{k - ½}{n}.$$

In Theorem 3 below we show the mean and variance of the order statistics for both sampling methods and their mean-squared-error in estimating the two quantile probabilities above. We use $\text{MSE}_{n,k}$ and $\text{MSE}_{n,k}^*$ to denote the mean-squared-error for $U_{(k)}$ and $U_{(k)}^*$ respectively, with the subsequent function input showing the stipulated quantile probability as the estimand.

**THEOREM 3:** The order statistics have mean and variance given by:

$$\mathbb{E}(U_{(k)}) = p_k \qquad \mathbb{V}(U_{(k)}) = \frac{p_k(1-p_k)}{n+2},$$

$$\mathbb{E}(U_{(k)}^*) = p_k^* \qquad \mathbb{V}(U_{(k)}^*) = \frac{1}{12n^2}.$$

Taken as estimators of $p_k$ they have the mean-squared error values:

$$\text{MSE}_{n,k}(p_k) \equiv \mathbb{E}((U_{(k)} - p_k)^2) = \frac{p_k(1-p_k)}{n+2},$$

$$\text{MSE}_{n,k}^*(p_k) \equiv \mathbb{E}((U_{(k)}^* - p_k)^2) = \frac{1}{3n^2} - \frac{p_k(1-p_k)}{n^2}.$$

Taken as estimators of $p_k^*$ they have the mean-squared error values:

$$\text{MSE}_{n,k}(p_k^*) \equiv \mathbb{E}((U_{(k)} - p_k^*)^2) = \frac{(n-2)p_k^*(1-p_k^*) + ¾}{(n+1)(n+2)},$$

$$\text{MSE}_{n,k}^*(p_k^*) \equiv \mathbb{E}((U_{(k)}^* - p_k^*)^2) = \frac{1}{12n^2}.$$

**COROLLARY:** Taking $\phi = k/n$ and $n \to \infty$ gives the asymptotic equivalence:

$$\text{MSE}_{n,k}(p_k) \approx \frac{\phi(1-\phi)}{n} \qquad \text{MSE}_{n,k}(p_k^*) \approx \frac{\phi(1-\phi)}{n},$$

$$\text{MSE}^*_{n,k}(p_k) \approx \frac{1-3\phi(1-\phi)}{3n^2} \qquad \text{MSE}^*_{n,k}(p_k^*) = \frac{1}{12n^2}.$$

which gives the related asymptotic equivalence:

$$\log \text{MSE}_{n,k}(p_k) - \log \text{MSE}_{n,k}(p_k) \approx \text{const} + \log n + r(\phi),$$

$$\log \text{MSE}_{n,k}(p_k^*) - \log \text{MSE}_{n,k}(p_k^*) \approx \text{const} + \log n + r_*(\phi),$$

where $r(\phi) \equiv \log(\phi(1-\phi)) - \log(1-3\phi(1-\phi))$ and $r_*(\phi) \equiv \log(\phi(1-\phi))$.

From Theorem 3 and its corollary we can see that there is generally a closer correspondence for the empirical quantiles from quantile-stratified sampling to the true quantiles than for the IID sampling. The order statistics from IID sampling give empirical quantiles that are unbiased estimators for $p_k$ and the order statistics from QS sampling give empirical quantiles that are unbiased estimators for $p_k^*$. Even allowing for bias, measured in terms of the mean-squared-error in estimation of the true quantiles, the empirical quantiles from IID sampling are order $\mathcal{O}(n^{-1})$ and the empirical quantiles from QS sampling are order $\mathcal{O}(n^{-2})$, so QS sampling gives a superior estimator for large $n$.

In Figures 3A-3B below we show bubble plots of the differences in log-MSE for estimation of the two types of quantiles over a matrix of values of $1 \leq k \leq n \leq 20$. The MSE for the two methods is equal when $n = 1$ and it is lower for QS sampling in all other cases.[4] This occurs even when estimating quantiles of the form $p_k$, for which IID sampling gives an unbiased estimator and QS sampling gives a biased estimator (but with lower variance). Moreover, the relative difference in accuracy of the methods becomes larger as the sample size $n$ increases. This is consistent with our findings of the order of the MSE approximation for each method shown in the corollary to Theorem 3 above. This establishes that the empirical quantiles from QS sampling adhere closer to the true quantiles than the empirical quantiles from IID sampling.

[4] In the case where $n = 1$ the IID and QS sampling methods are identical, so the equality of MSE in this case is a necessary result of this.

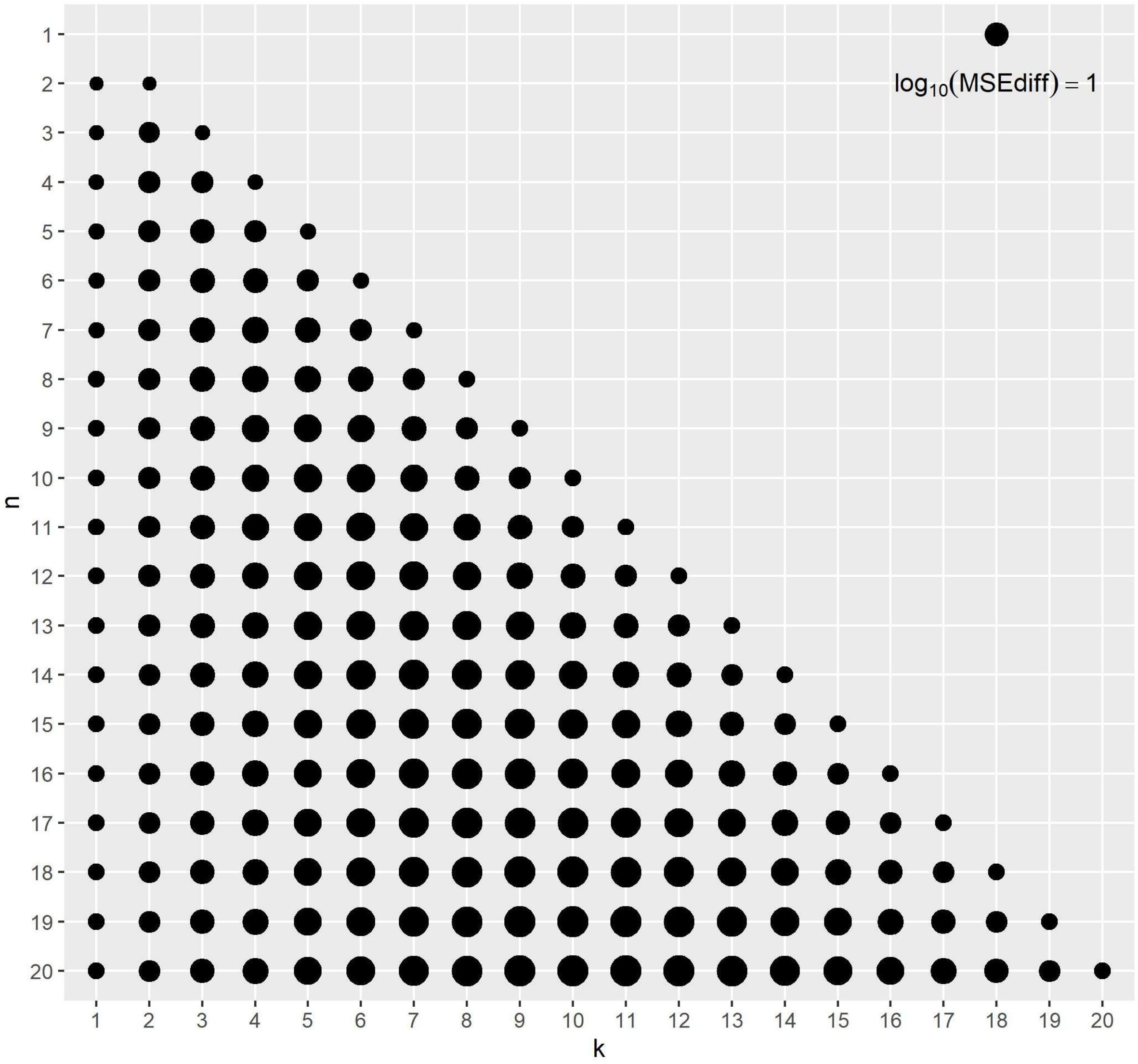

**FIGURE 3A:** Difference in log-mean-squared-error for estimating $p_k$
(all cases have lower MSE for QS sampling)

The present results pertaining to the order statistics elucidate why we saw greater regularity in the QQ plots for QS sampling than for IID sampling in Figures 1-2. The empirical quantiles from QS sampling adhere much more closely to the true quantiles than IID sampling, so the empirical distribution of the sample is closer to the sampling distribution. This is a reflection of the fact that QS sampling is analogous to "sampling without replacement" for an arbitrary distribution, so it has more complete coverage of the true sampling distribution. This is the primary benefit of QS sampling as an alternative to IID sampling for the generation of sample values from a distribution.

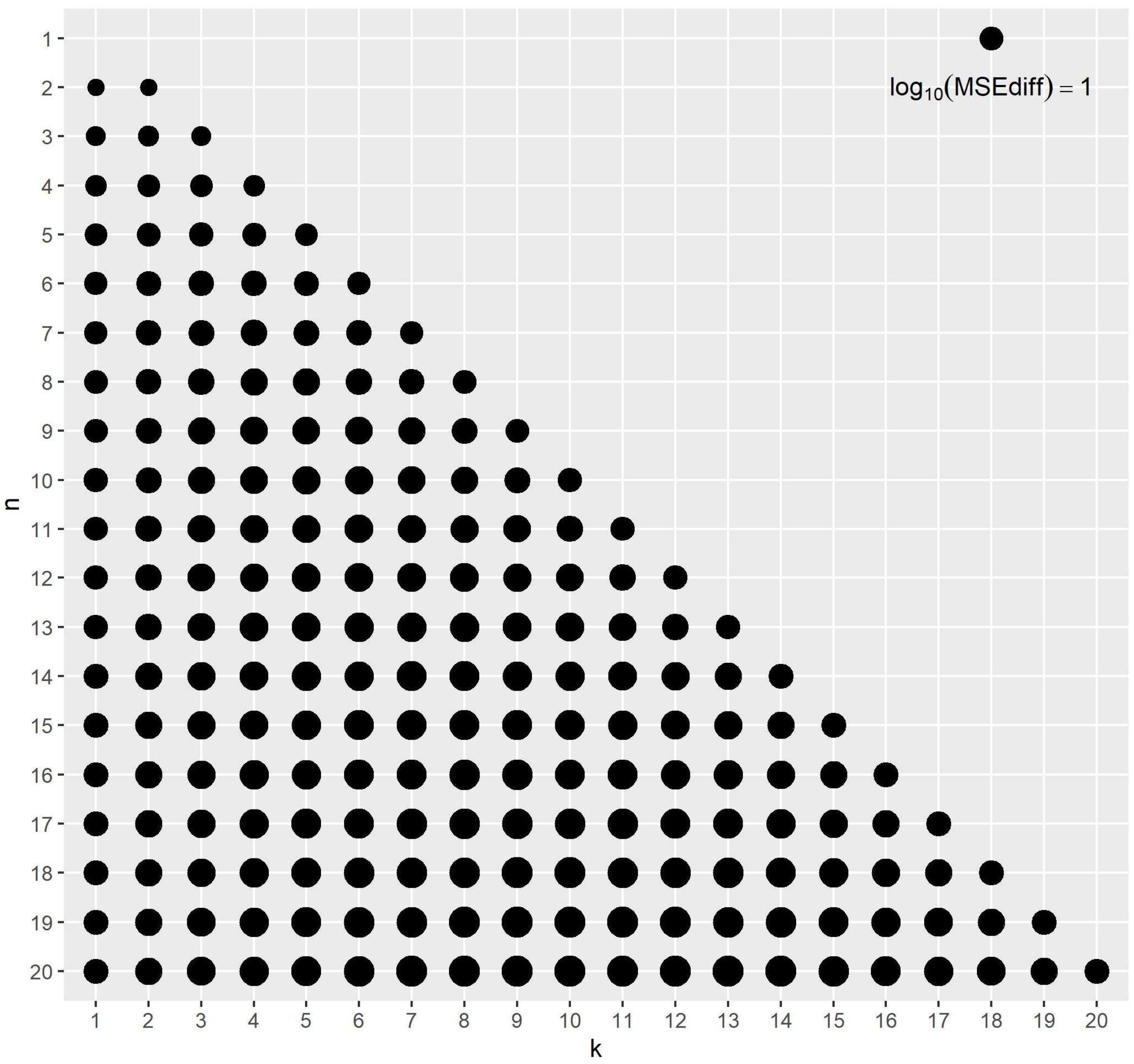

**FIGURE 3B:** Difference in log-mean-squared-error for estimating $p_k^*$
(all cases have lower MSE for QS sampling)

Having established the accuracy of the order statistics as estimators of the true quantiles of the sampling distribution, it is also useful to look at the regularity of the spacing between the order statistics from each sampling method. For this purpose, we define the differences between order statistics of the uniform sample values by:

$$D_{k,\ell} \equiv U_{(k+\ell)} - U_{(k)} \qquad 1 \leq k < k + \ell \leq n,$$

$$D_{k,\ell}^* \equiv U_{(k+\ell)}^* - U_{(k)}^* \qquad 1 \leq k < k + \ell \leq n.$$

In Theorem 4 and its corollary below, we show that QS sampling gives more consistency in spacing between order statistics. This is reflected in the lower variance for the spacing between order statistics from QS uniform random variables.

**THEOREM 4:** The differences in order statistics have the following distributions:

$$D_{k,\ell} \sim \text{Beta}(\ell, n-\ell+1),$$

$$D^*_{k,\ell} \sim \text{Triangular}\left(\frac{\ell-1}{n}, \frac{\ell}{n}, \frac{\ell+1}{n}\right).$$

These differences have mean and variance given by:

$$\mathbb{E}(D_{k,\ell}) = \frac{\ell}{n+1} \qquad \mathbb{V}(D_{k,\ell}) = \frac{\ell(n-\ell+1)}{(n+1)^2(n+2)},$$

$$\mathbb{E}(D^*_{k,\ell}) = \frac{\ell}{n} \qquad \mathbb{V}(D^*_{k,\ell}) = \frac{1}{6n^2}.$$

**COROLLARY:** Taking $\phi = k/n$, $\psi = \ell/n$ and $n \to \infty$ gives the asymptotic equivalence:

$$\mathbb{E}(D_{k,\ell}) \approx \psi \qquad \mathbb{V}(D_{k,\ell}) \approx \frac{\psi(1-\psi)}{n},$$

$$\mathbb{E}(D^*_{k,\ell}) \approx \psi \qquad \mathbb{V}(D^*_{k,\ell}) = \frac{1}{6n^2}.$$

From Theorem 4 and its corollary we can see that there is generally greater regularity in the spacing between order statistics from QS sampling than from IID sampling. The variance of the spacing between order statistics from IID sampling are order $\mathcal{O}(n^{-1})$ and the variance of the spacing between order statistics from QS sampling are order $\mathcal{O}(n^{-2})$, so QS sampling gives significantly more regular spacing for large $n$. This means that QR sampling is quadratically faster than IID sampling for estimating quantiles of a distribution; this performance is similar to the quadratically faster convergence for estimation of linear quantities for Latin hypercube sampling (see e.g., McKay, Beckman and Conover 1979; Stein 1987; Owen 1992). Taking Theorems 3-4 in conjunction, we see that compared to IID sampling, QS sampling has greater overall adherence and regularity of the empirical quantiles to the true quantiles of the sampling distribution. Our analysis here has not been extended to LQS sampling (since the formulae at issue become very cumbersome and complicated with multiple layers) but this case operates part-way between the pure QS sample and the IID sample. Non-reductive versions of LQS sampling (i.e., those that don't reduce to pure QS sampling or IID sampling) have greater adherence/regularity of the empirical quantiles to the true quantiles of the sampling distribution than IID sampling, but less than QS sampling.

## 5. Generating quasi-IID and quasi-QS samples using MCMC samples

The standard method for generating pseudo-random numbers from a univariate probability distribution is to generate pseudo-random uniform values and then use inverse transformation sampling by applying the quantile function of the distribution to transform the uniform values to values from the stipulated distribution. This is a useful method in cases where the quantile function is easily computed, but it is less useful in cases where the quantile function is difficult to compute.

In this section we will consider the case of generating a random sample where the only available probability function is the kernel of the density function —i.e., a function $K(x) \propto f(x)$. When only the kernel function is available, it is possible to use Markov Chain Monte Carlo (MCMC) methods such as the Metropolis-Hastings algorithm to generate a sequence of non-independent (autocorrelated) values from a Markov chain with stationary distribution equal to the stipulated distribution (see e.g., Brooks et al. 2011, Hanada and Matsuura 2022). Typically, this is done by generating a long chain of values and removing a sufficient number of "burn-in" values to ensure that the Markov chain has converged close to its stationary distribution. Values from the chain are autocorrelated, but the correlation diminishes as the values become further apart in the chain, allowing the user to obtain a quasi-IID sample from the distribution by sampling at intervals which are sufficiently far apart in the chain to have near-zero correlation.

Here we will consider how to obtain a quasi-IID or quasi-QS sample using a chain of values generated using MCMC methods. To do this, suppose we set some "multiplier" value $r \in \mathbb{N}$ and generate a set of values $M_1, \dots, M_{nr}$ from an appropriate MCMC method with stationary distribution equal to the stipulated sampling distribution of interest. (This could be a chain of values generated by the Metropolis-Hastings algorithm or some other MCMC method; we will be agnostic here, but we will make some assumptions about the behaviour of this chain later.) We will assume that this sample already excludes a sufficient number of "burn-in" values from that we can consider it to be marginally distributed according to the stipulated distribution of interest. We will also assume that the sample satisfies relevant diagnostic criteria for "good coverage" of the sampling distribution (e.g., based on inspection of trace plots, autocorrelation plots, etc.). The order statistics from the generated sample (which may include duplicates) can be denoted as $M_{(1)} \leq M_{(2)} \leq \cdots \leq M_{(nr)}$. If the multiplier $r$ is sufficiently large (giving a large

"effective sample size" for the sample) then the conditional distribution for the stipulated sampling distribution over each quantile-block $s = 1, \dots, n$ can be approximated reasonably well by a discrete uniform distribution over the $r$ values:

$$M_{((s-1)r+1)}, \dots, M_{((s-1)r+r)} \qquad s = 1, \dots, n.$$

Consequently, we may generate a quasi-IID sample or a quasi-QS sample as follows:

$$T_1, \dots, T_n \sim \text{SRSWR}\{1, \dots, r\}$$

$$\text{Quasi-IID:} \qquad S_1, \dots, S_n \sim \text{SRSWR}\{1, \dots, n\} \qquad X_i \equiv M_{((S_i-1)r+T_i)},$$

$$\text{Quasi-QS:} \qquad S_1^*, \dots, S_n^* \sim \text{SRSWOR}\{1, \dots, n\} \qquad X_i^* \equiv M_{((S_i^*-1)r+T_i)}.$$

Let $f$ and $F$ denote the density function and distribution function of the sampling distribution (i.e., the stationary distribution of the Markov chain). We will refer to these processes using the following simple shorthand:

$$X_1, \dots, X_n \sim \text{Quasi-IID } f,$$

$$X_1^*, \dots, X_n^* \sim \text{Quasi-QS } f.$$

**REMARK:** The above method for producing a quasi-IID sample is just one possible method, which we have tailored to be analogous to the quasi-QS sample. There are other acceptable variants of the above method that may be used to generate a quasi-IID sample. One acceptable variant is to instead sample from the original (non-ordered) chain of values at fixed intervals of $r$ values, yielding quasi-IID values with low correlation when $r$ is not too small. Another variant is to perform the quasi-IID sample in a way that does not allow supplicate drawing of the same value in the MCMC chain. □

The quasi-QS sampling method shown above is one that can be implemented for any univariate distribution where it is possible to establish an MCMC chain with stationary distribution given by the stipulated sampling distribution; it does not require a known quantile function. It works well when there is a large effective sample size for the MCMC chain.[5] Although it is highly general, the method requires much more computation than direct QS sampling using a quantile function, because there is a substantial amount of computation required to set up and execute the MCMC chain (and run relevant diagnostics to ensure convergence) and then only one in every $r$ values in the chain is actually used for the quasi-QS sample. Moreover, sorting to get the order statistics from the MCMC chain incurs a computational cost of $\mathcal{O}(nr\log(nr))$, so computation can become burdensome compared to QS sampling when $r$ is large.

[5] It can be useful to augment this method by using the sample autocorrelation function for the MCMC chain to estimate the "effective sample size" of the Markov chain. If the effective sample size is large then it is likely that the ordered MCMC values adhere closely to the true quantile function for the sampling distribution, so the quasi-QS sampling method will then closely approximate a true QS-sample from the underlying sampling distribution.

If we are willing to stipulate that the multiplier $r$ is large enough to give a large effective sample size —such that the order statistics from the MCMC chain follow the standard distribution for the order statistics from an IID sample— then we may obtain results for the accuracy of estimation of the true quantiles in the sampling distribution that are similar to our previous results for the QS sample. In Theorem 5 below we take $X_1^*, \dots, X_n^* \sim \text{Quasi-QS } f$ and we show the moments of the uniform order statistics from this sample, as well as the mean-squared-error of these uniform order statistics in estimating the true quantiles $p_1^*, \dots, p_n^*$. It is important to note that the uniform order statistics are unobserved since we do not assume knowledge of $F$ in the quasi-QS sampling method.

**THEOREM 5:** Suppose that $r$ is sufficiently large so that the values $M_{(1)} \leq M_{(2)} \leq \cdots \leq M_{(nr)}$ follow the standard distribution for the order statistics from an IID sample, and consider the quasi-QS sample $X_1^*, \dots, X_n^* \sim \text{Quasi-QS } f$. Letting $U_{(k)}^* = F(X_{(k)}^*)$ denote the order statistics of the uniform values from the sample, these order statistics have mean and variance given by:

$$\mathbb{E}(U_{(k)}^*) = \frac{(k - ½)r + ½}{nr + 1},$$

$$\mathbb{V}(U_{(k)}^*) = \frac{nr^3 + [12n(k - ½) - 12k(k - 1) - 2]r^2 + 5nr + 2}{12(nr + 1)^2(nr + 2)}.$$

Taken as an estimator of $p_k^*$ the order statistics have the mean-squared error values:

$$\text{MSE}_{n,r,k}^*(p_k^*) = \frac{n^2r^2 + 12n(k - ½)(n - k + ½)r + 8n^2 - 24(k - ½)n + 24(k - ½)^2}{12n^2(nr + 1)(nr + 2)}$$

**COROLLARY:** Taking $\phi = k/n$ and $n \to \infty$ gives the asymptotic equivalence:

$$\text{MSE}_{n,r,k}^*(p_k^*) \approx \frac{\phi(1 - \phi)}{nr}.$$

**COROLLARY:** As $r \to \infty$ we obtain $\mathbb{E}(U_{(k)}^*) \to p_k^*$ and $\text{MSE}_{n,k}^*(p_k^*) \to \mathbb{V}(U_{(k)}^*) \to 1/12n^2$.

As can be seen, this is a similar form to the moment/MSE results in Theorem 3 for the standard QS sample. In particular, the asymptotic MSE of the quasi-QS sample values as estimators of the corresponding quantiles of the distribution has the same form as in QS sampling. Broadly speaking, this means that we can generate a reasonable approximation to a QS sample without knowledge of the quantile function of the distribution, using a kernel of the density to generate an MCMC sample. This also imposes a computational burden, so it is worth considering the alternative of computing the quantile function directly from the kernel function, even if the latter conversion is also computationally burdensome.

The moment results in Theorem 5 are predicated on an approximation to the true distribution of the order statistics, which assumes that the sample size is large enough to effectively ignore the autocorrelation in the MCMC process. In cases where the MCMC process is sufficiently well described and understood, it may be is possible to undertake a more accurate analysis by looking at the distribution of the order statistics from autocorrelated processes (see Serinaldi, Lombardo and Kilsby 2020). This requires specification of the nature of the autocorrelation in the process, and it involves distributional and moment results that are highly complicated and cumbersome, so the value of this exercise will typically be outweighed in practice by just using that time to undertake the simulation with a higher value of $r$.

Because the MCMC algorithm can generate duplicate values (e.g., due to an accept/reject step for a candidate value), there is a possibility that these sampling methods can contain duplicate values. This is significantly less likely (but still possible) for the quasi-QS sample, because the former samples from the quantile-blocks without repetition. Duplicates in this latter method can only occur when there are duplicate values that span consecutive quantile-blocks, and they are each selected as the values sampled from their respective quantile blocks. It is possible to assess the adequacy of the multiplier $r$ by looking at the probability of generating duplicate values in the quasi-QS sample depending on the multiplier. This is one possible criterion by which one may determine the appropriate multiplier for the method. Again, the practical value of such work will typically be outweighed in practice by just using that time to undertake the simulation with a higher value of $r$.

## 6. Estimation of mean quantities using quantile-stratified simulation

Simulation methods can be employed to estimate quantities that can be expressed as expected values of a function of a random variable. Such quantities are typically integrals of a function over a sample space, expressed in a form that can be decomposed into a density function for a random variable and a remaining term acting as a function on the outcome of that random variable. Deterministic methods may be used to compute such integrals, but simulation-based methods using random simulations offer a useful alternative. Importance sampling is the most common type of simulation method for this problem, but there are a broad class of Monte Carlo methods that can be employed.

Suppose we have a univariate function $H: \mathbb{R} \to \mathbb{R}$ and we want to estimate the expected value:

$$\mu \equiv \mathbb{E}(H(X)) \qquad X \sim f.$$

We can estimate this quantity by generating random samples from the sampling distribution (with density $f$) and using the sample mean of the resulting sample as an estimator. We will consider two variants of this process using IID sampling and QS sampling, with sample means:

$$\hat{\mu}_n \equiv \frac{1}{n}\sum_{i=1}^n H(X_i) \qquad \hat{\mu}_n^* \equiv \frac{1}{n}\sum_{i=1}^n H(X_i^*) .$$

The values of the function can be related to the underlying uniform values from each of these simulation methods via the fact that:

$$H(X_i) = H(Q(U_i)) = H \circ Q(U_i),$$
$$H(X_i^*) = H(Q(U_i^*)) = H \circ Q(U_i^*).$$

Taking $G = H \circ Q$ to be the composition function we then have the alternative form:

$$\hat{\mu}_n = \frac{1}{n}\sum_{i=1}^n G(U_i) \qquad \hat{\mu}_n^* = \frac{1}{n}\sum_{i=1}^n G(U_i^*) .$$

It is simple to show that both estimators are unbiased, which means that a reasonable way to assess their relative merits is by looking at their variances. Here is where things get a bit murky and depend on the particular forms of the functions, but we can still make some observations at an approximate level. In particular, if the composition $G$ is a sufficiently "well behaved" function then the negative correlation between the quantile-stratified uniform random variables may follow through to the quantile-stratified values of the function at issue, giving the negative correlation $\mathbb{C}(G(U_i^*), G(U_j^*)) < 0$. In this case we expect the variance of the estimator using QS sampling to be lower than the variance of the estimator using IID sampling.

This heuristic reasoning is bolstered by noting the fact that a first-order Taylor approximation to any function is a linear approximation, which makes it possible to derive approximations for the cases of interest. The first-order Taylor approximations to the variance and covariance of the transformed uniform random variables are:

$$\mathbb{V}(G(U_i)) \approx G'(\mathbb{E}(U_i))^2 \cdot \mathbb{V}(U_i) = \frac{G'(½)^2}{12},$$

$$\mathbb{C}(G(U_i), G(U_j)) \approx G'(\mathbb{E}(U_i))G'(\mathbb{E}(U_j))\,\mathbb{C}(U_i, U_j) = 0,$$

$$\mathbb{V}(G(U_i^*)) \approx G'(\mathbb{E}(U_i^*))^2 \cdot \mathbb{V}(U_i^*) = \frac{G'(½)^2}{12},$$

$$\mathbb{C}(G(U_i^*), G(U_j^*)) \approx G'(\mathbb{E}(U_i^*))G'(\mathbb{E}(U_j^*))\,\mathbb{C}(U_i^*, U_j^*) = -\frac{G'(½)^2(n+1)}{12n^2}.$$

This first-order approximation preserves the negative correlation between the values (except in the case where $G'(½) = 0$) and gives the approximate estimator variances:

$$\begin{aligned}
\mathbb{V}(\hat{\mu}_n) &= \frac{1}{n^2}\left[\sum_{i=1}^{n} \mathbb{V}(G(U_i)) + \sum_{i \neq j} \mathbb{C}(G(U_i), G(U_j))\right] \\
&\approx \frac{1}{n^2}\sum_{i=1}^{m} \frac{G'(½)^2}{12} \\
&= \frac{G'(½)^2}{12n},
\end{aligned}$$

$$\begin{aligned}
\mathbb{V}(\hat{\mu}_n^*) &= \frac{1}{n^2}\left[\sum_{i=1}^{n} \mathbb{V}(G(U_i^*)) + \sum_{i \neq j} \mathbb{C}(G(U_i^*), G(U_j^*))\right] \\
&\approx \frac{1}{n^2}\left[\sum_{i=1}^{n} \frac{G'(½)^2}{12} - \sum_{i \neq j} \frac{G'(½)^2(n+1)}{12n^2}\right] \\
&= \frac{G'(½)^2}{n}\left[\frac{1}{12} - \frac{(n+1)(n-1)}{12n^2}\right] \\
&= \frac{G'(½)^2}{12n}\left[1 - \frac{n^2-1}{n^2}\right] \\
&= \frac{G'(½)^2}{12n} \cdot \frac{1}{n^2} \\
&= \frac{G'(½)^2}{12n^3}.
\end{aligned}$$

Under these variance approximations, estimation using IID sampling gives a standard deviation that is $\mathcal{O}(n)$ larger than for QS sampling. This is only a crude approximation based on Taylor approximation to the relevant moments, but they correspond to known asymptotic results for Latin hypercube sampling for estimation of well-behaved functions (see e.g., McKay, Beckman and Conover 1979; Stein 1987; Owen 1992). In an analysis of Latin hypercube sampling, Owen (1992) established an $\mathcal{O}(n^{-3})$ variance result analogous to the above for any function that is continuously differentiable with bounded first derivative. Because QS sampling is equivalent to randomised one-dimensional Latin hypercube sampling, asymptotic results from this field apply to QS sampling and give broad support for the intuition of lower variance for estimation of moment quantities. For more general functions that are not well-behaved (e.g., non-differentiable, steep changes, etc.) the variance from QS sampling may be no better than from IID sampling. Indeed, it is possible that the form of the function $G$ will be such that the negative correlation between the underlying uniform values in the QS sample is "flipped" to positive correlation between the transformed values, in which case the estimator using QS sampling may actually be worse than the estimator using IID sampling for functions that are extremely poorly behaved.

Whilst it is possible for QS sampling to perform worse than IID sampling for extremely poorly behaved functions, the broad asymptotic theory of Latin hypercube sampling provides reason to believe that the method will generally reduce the variance of moment estimators. In general, the preservation of negative correlation between values in QS sampling is more common for a wide class of functions than flipping to positive correlation, and this reasoning also gives us a general sense that estimation based on QS sampling will generally be superior to estimation based on IID sampling in a wide class of cases.

Because QS sampling adheres more closely to the quantiles of the true sampling distribution than IID sampling, estimation of mean quantities using QS sampling can be regarded as a halfway point between estimation of mean quantities using IID simulation and estimation using deterministic methods based on points spaced over the range of the integral (e.g., Simpson's method and variants thereof). Using QS sampling preserves the benefits of stochastic methods that use non-deterministic points in the support of the distribution, while (usually) lowering the variance of the estimator compared to using an IID sample.

**Application to importance sampling:** One plausible application of QS sampling is in the use of importance sampling to estimate an integral representing an expected value of a function of a random variable. Importance sampling involves generating a random sample from a proposal distribution (usually continuous) and taking a weighted average of a function of the outcomes as an estimator of the expected value at issue. In cases where the proposal distribution and the function used for estimation are both continuous and "well behaved" (in the sense previously discussed) it is plausible that greater adherence to the true quantiles of the proposal distribution could potentially improve estimation, meaning that QS sampling may be a useful alternative to IID sampling in this case.

To estimate the true mean quantity $\mu$ using **importance sampling**, we let $g$ be an alternative proposal density with a known and computable quantile function (making it simple to simulate random variables with this density for a QS or IID sample). Using this proposal density, we write the integral of interest in alternative form as:

$$\mu = \int_{\mathbb{R}} H_{\bullet}(x) g(x) \, dx \qquad H_{\bullet}(x) \equiv \frac{H(x)f(x)}{g(x)}.$$

In **standard importance sampling** we simulate $X_1, \dots, X_n \sim \text{IID}\, g$ and we then approximate the integral of interest by the sample moment:

$$\hat{\mu}_n \equiv \frac{1}{n} \sum_{i=1}^{n} H_{\bullet}(x_i) = \frac{1}{n} \sum_{i=1}^{n} \frac{H(x_i)f(x_i)}{g(x_i)}.$$

In **quantile-stratified importance sampling** we instead simulate $X_1^*, \dots, X_n^* \sim \text{QS}\, g$ and we then approximate the integral of interest by the sample moment:

$$\hat{\mu}_n^* \equiv \frac{1}{n} \sum_{i=1}^{n} H_{\bullet}(x_i^*) = \frac{1}{n} \sum_{i=1}^{n} \frac{H(x_i^*)f(x_i^*)}{g(x_i^*)}.$$

It is simple to establish that $\hat{\mu}_n$ and $\hat{\mu}_n^*$ are unbiased estimator of $\mu$. Taking $G_{\bullet} = H_{\bullet} \circ Q$ to be the composition function acting on the underlying uniform random variables, the variance of the latter estimator will be lower in the case where the transformation $G_{\bullet}$ preserves negative correlation between the values. The variance of the estimator in standard importance sampling is minimised when $g(x) \propto |H(x)| f(x)$ (which we demonstrate below) so the method works well when using a proposal density that is close to this proportionality requirement. Things are slightly more complicated for the quantile-stratified importance sampling (owing to correlation between the QS sample values), but this form of the proposal density should still give a good estimator in this latter case.

**EXAMPLE A (Importance sampling using the beta distribution):** Suppose we wish to use simulation to estimate the quantity:

$$\mu \equiv \int_0^1 x \ln(x) \, \text{Beta}(x|2,2) \, dx = -0.2916667.$$

(We have shown the true value of the integral, which is computed using a formula involving the digamma function.) Using the proposal density $g(x) = \text{Beta}(x|3,2)$ we can write this integral in the alternative form:

$$\mu = \int_0^1 H_{\bullet}(x) \, g(x) \, dx \, ,$$

using the importance function:

$$H_{\bullet}(x) = \frac{H(x) f(x)}{g(x)} = \frac{x \ln(x) \, \text{Beta}(x|2,2)}{\text{Beta}(x|3,2)} = \frac{\Gamma(3)\,\Gamma(4)}{\Gamma(2)\,\Gamma(5)} \cdot \ln(x) \, .$$

As an illustration of both methods of important sampling, we generate importance samples to estimate the integral of interest using both IID and QS sampling using $n = 100$ values. We generate one-thousand simulations from each method and give violin plots of these estimates in Figure 4A below. The standard errors and root-mean-squared-errors of the estimators in these simulations are also shown here:

| | IID | QS |
|---|---|---|
| StdErr | 0.02070450 | 0.00176950 |
| RMSE | 0.02069569 | 0.00176862 |

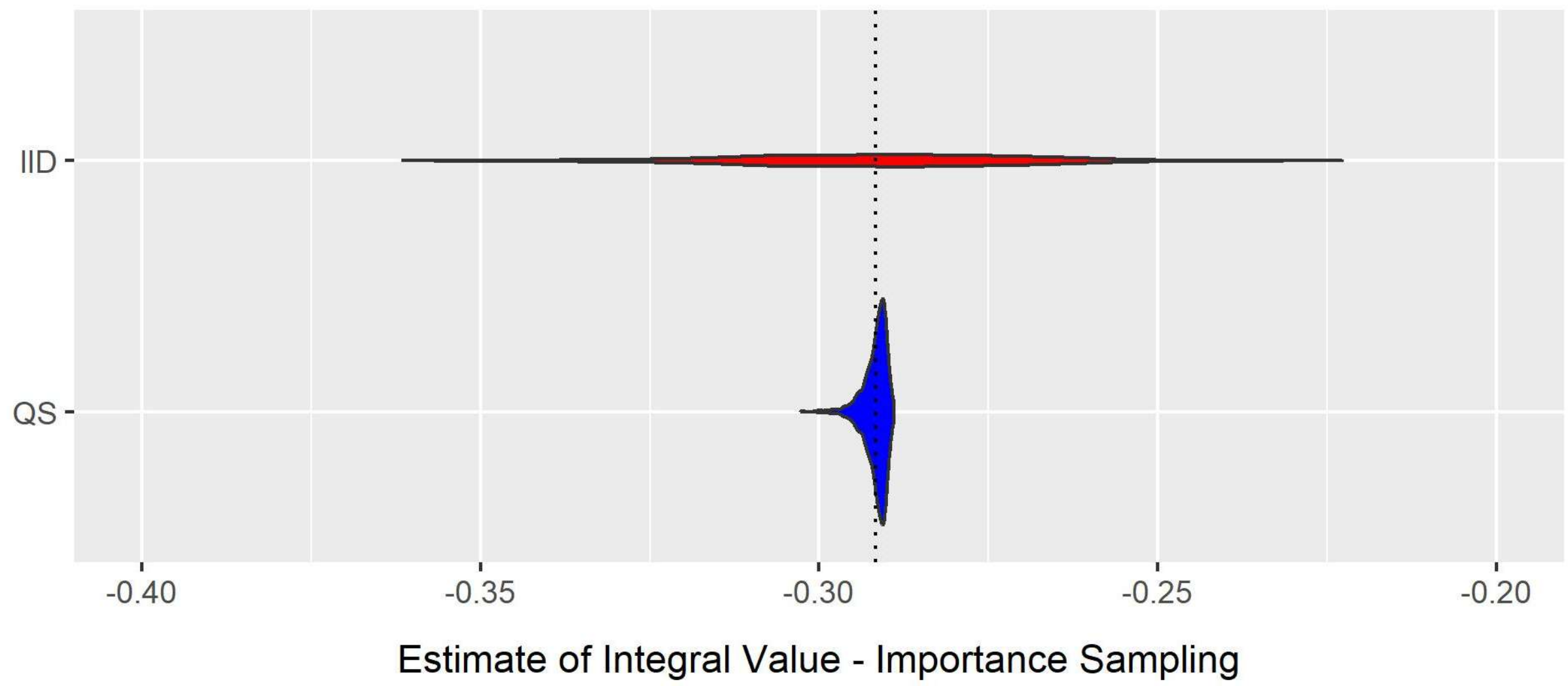

**FIGURE 4A:** Violin plots for one-thousand simulations of importance sampling estimate (QS sampling in blue— IID sampling in red — true value is vertical line)

**EXAMPLE B (Importance sampling using the gamma distribution):** Suppose we wish to use simulation to estimate the quantity:

$$\mu \equiv \int_0^\infty \exp(-x^2)\,\text{Ga}(x|2,5)\,dx = 0.8236078.$$

(We have shown the true value of the integral.) Using the proposal density $g(x) = \text{Ga}(x|2,6)$ we can write this integral in the alternative form:

$$\mu = \int_0^1 H_\bullet(x)\,g(x)\,dx\,,$$

using the importance function:

$$H_\bullet(x) = \frac{H(x)f(x)}{g(x)} = \frac{\exp(-x^2)\,\text{Ga}(x|2,5)}{\text{Ga}(x|2,6)} = \frac{5^2\cdot\Gamma(6)}{6^2\cdot\Gamma(5)}\cdot\exp(x(1-x))\,.$$

As an illustration of both methods of important sampling, we generate importance samples to estimate the integral of interest using both IID and QS sampling using $n = 100$ values. We generate one-thousand simulations from each method and give violin plots of these estimates in Figure 4B below. The standard errors and root-mean-squared-errors of the estimators in these simulations are also shown here:

| | IID | QS |
|---|---|---|
| StdErr | 0.006269564 | 0.001279065 |
| RMSE | 0.006266765 | 0.001278935 |

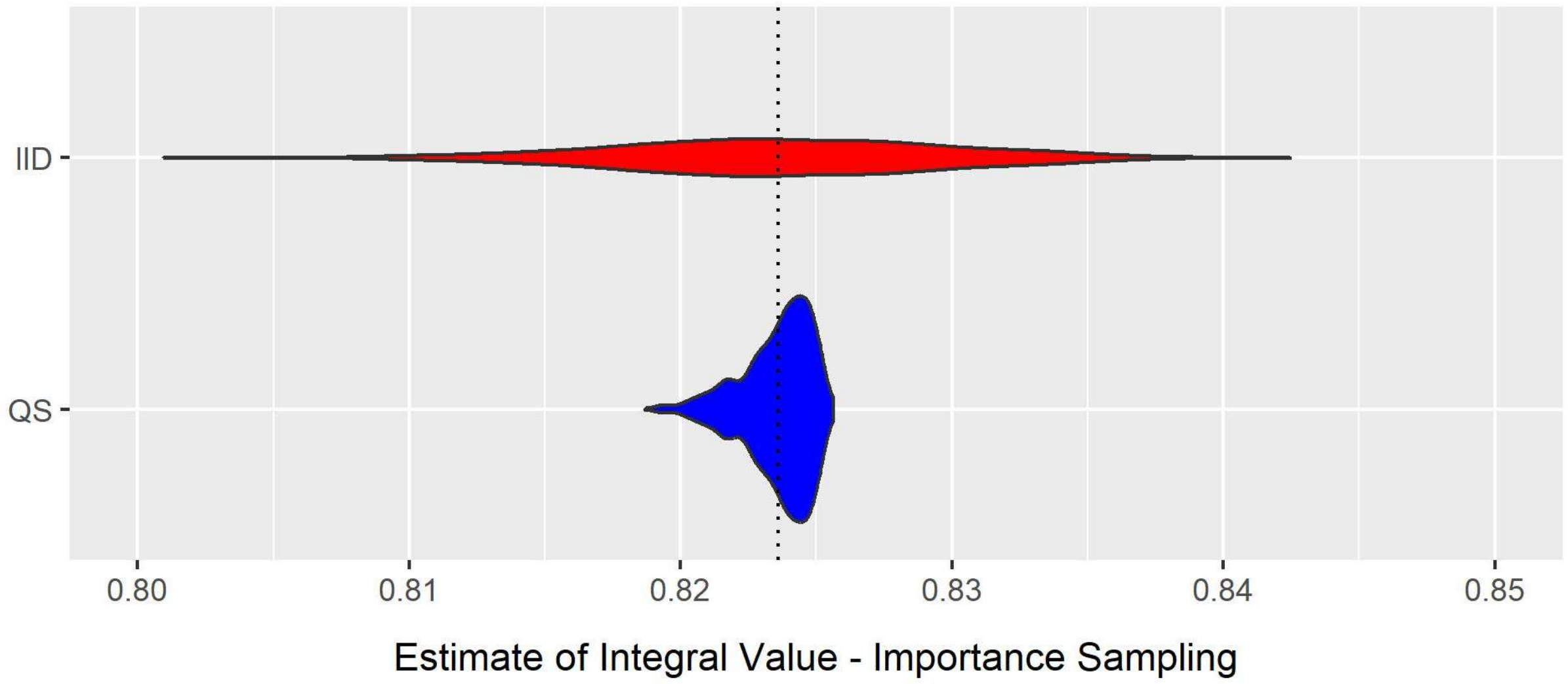

**FIGURE 4B:** Violin plots for one-thousand simulations of importance sampling estimate (QS sampling in blue— IID sampling in red — true value is vertical line)

The above examples illustrate the use of the QS importance sampling method, which operates by using a QS sample from the proposal distribution instead of an IID sample. In the examples shown, the estimates from QS importance sampling are significantly more accurate than the estimates from standard importance sampling. This occurs because the QS sample gives a more stable adherence to the true quantiles of the proposal distribution, with less variability in the empirical quantiles from the sample. Another way of looking at the difference is that the QS sample gives negatively correlated values from the proposal distribution and this reduces the variance of the resulting mean estimator. Improvement of importance sampling is just one possible application of QS sampling in making estimates of the mean of a function of a random variable, but it is quite a broad area of application.

## 7. Simulation analysis of accuracy of QS importance sampling

The heuristic reasoning and examples in the section above show that QS sampling is effective for importance sampling and it typically yields more accurate estimates than IID sampling with the same sample sizes. The degree to which this increase in accuracy obtains is complicated and depends on several input factors, including the nature of the target distribution and proposal distribution, the distance between these distributions, and the diffuseness of the distributions. Although we obtained some rough accuracy results for the importance sampling estimators in the previous section, these were based on a first-order Taylor approximation, which is only a crude approximation. In this section we conduct a simulation analysis to determine the relative accuracy of importance sampling estimation using IID and QS samples over various simulation problems involving different target and proposal distributions and different sample sizes.

For our simulation analysis we have chosen to use $D = 20$ different continuous distributions with support on the positive real numbers and with means and standard deviations that vary by small amounts, but not so much that the distributions will give wildly different values. This is consistent with the general approach of importance sampling where the proposal distribution is generally chosen to be close to the kernel of the target. The distributions used here, and their corresponding means and standard deviations, are shown in Appendix 2, along with details of the simulation. These distributions were chosen because they represent a broad class of cases from commonly used families on a common support. While the distributions vary roughly in shape, they are all unimodal distributions with support on the positive real numbers, so the present analysis should be taken as looking at distributions that are "reasonably similar" but with slight variation in shape and moments. This approach allows us to determine the accuracy of IID vs SQ sampling for importance sampling problems taken over a class of distributions that vary in their shape, mean, and variance.

Our simulation involves using importance sampling with a proposal distribution to estimate the mean of the target distribution. The simulation uses each ordered pair of these distributions as the proposal and target distributions. For each ordered pair of these distributions, we computed the Kullback-Leibler divergence using the proposal distribution as the base distribution, and we then simulated importance sampling estimates using IID samples and QS samples with sample sizes $n = 10, 30, 100, 300, 1000$. For each estimate we computed the log-root-mean-squared-error (LRMSE) against the known mean of the target distribution (using one-thousand simulations for this computation in each case). This gave us a simulation dataset composed of $20^2 \times 5 \times 2 = 4000$ data points, each giving the LRMSE from importance sampling using a particular ordered pair of distributions, sample size, and sampling method.

Using this dataset, we formed a regression model to predict the LRMSE of the estimator based on plausible explanatory variables. As a starting point for the model, we considered the case of importance sampling where the proposal and target distributions are identical with common mean $\mu$ and standard deviation $\sigma$, and the sample is drawn using IID sampling. In this special case, the LRMSE for the importance sampling estimator is $\text{LRMSE}_0 = \log(\sigma/\sqrt{n})$, which we will take as an offset term in the regression model. Let $\text{LRMSE}(i, j, n, \text{QS})$ denote the observed LRMSE from estimation using proposal distribution $i$, target distribution $j$, Kullback-Leibler divergence $KL_{i,j}$, sample size $n$, and QS as indicator of a QS sample. Our regression model has the form:[6]

$$\text{LRMSE}(i, j, n, \text{QS}) = \log\left(\frac{\sigma_i}{\sqrt{n}}\right) + \beta_0 + \beta_1 \log\left(\frac{\sigma_j}{\sigma_i}\right) + \beta_3 \log(KL_{i,j}) \\ + \beta_4 \log(n) + \beta_5 \text{QS} \log(n) + \text{Error}.$$

The offset term $\log(\sigma_i/\sqrt{n})$ used in this model represents the standard error obtained from IID sampling of the mean of the proposal distribution, which is the LRMSE obtained in the special case where the target distribution matches the proposal distribution, with IID sampling used as the sampling method. Consequently, the remaining terms in the model represent the deviation in accuracy from this base case. Given the nature of importance sampling, we would expect a priori that high value for $\log(\sigma_j/\sigma_i)$ or $KL_{i,j}$ would make estimation less accurate and we also expect that QS sampling would make estimation more accurate. We fit a Gaussian regression model with the above form to our simulation dataset and examined the relevant diagnostic plots for the residuals to determine whether there were any substantial deviations from the assumed model form (see Appendix 2 for details). Our diagnostic analysis showed some departure from linearity and evidence of positive skew and a heavy right-tail in the residuals. The explanatory variables explain 64.53% of the variation in the LRMSE.

[6] Note: This model omits a base term for QS but includes the interaction term for $\text{QS}\log(n)$. The reason for this is to enforce the model constraint that QS sampling and IID sampling be treated the same when $n = 1$ (since QS sampling is equivalent to IID sampling in this case). This constraint is enforced in the model even though our simulation data does not cover such a low sample size. As a check on this constraint, we also ran an alternative regression model with the inclusion of this base term; there was no statistically significant evidence for inclusion of this term (i.e., no evidence of a relationship with LRMSE conditional on the other terms in the model).

As we expected, the log-difference in standard-deviations and the Kullback-Leibler divergence are both found to have positive effects on the LRMSE of the importance sampling estimator. Conditional on the other explanatory variables, we find the QS sampling leads to lower LRMSE than IID sampling at all sample sizes, with the relative accuracy of QS sampling improving with higher sample sizes. This is exhibited in Figure 5 below, where we show the ratio of the multiplicative effects of the sampling type combined with the sample size on the LRMSE (holding other explanatory variables constant). With a sample size of $n = 10$ the QS sampling has a LRMSE that is 45.3% as large as for IID sampling. With a sample size of $n = 1000$ the QS sampling has a LRMSE that is 9.3% as large as for IID sampling.

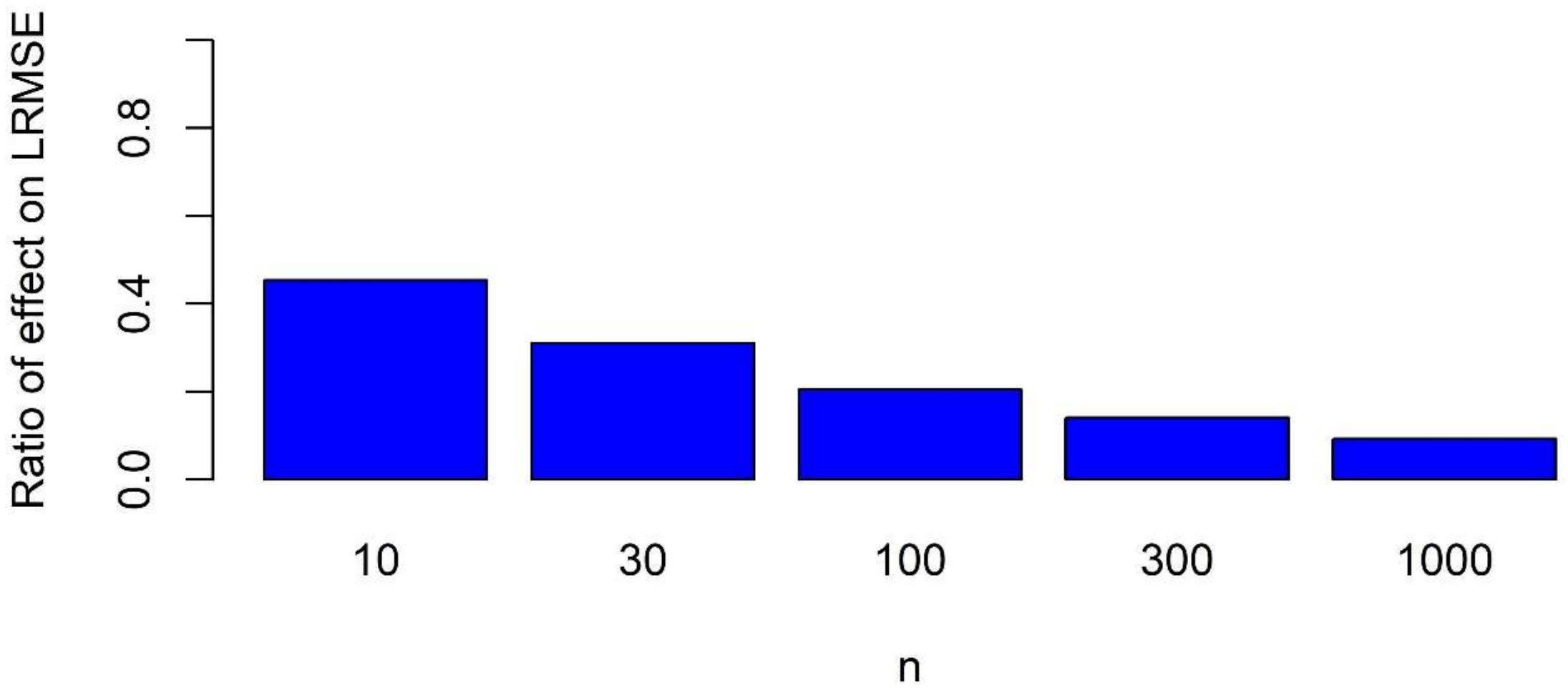

**FIGURE 5:** Ratio of multiplicative effects of QS sampling vs IID sampling on LRMSE (bars show QS effect/IID effect for each sample size; ratios less than one show that QS sampling is performing better than IID sampling)

The results of our simulation analysis provide reasonable explanatory power for the overall variation in the LRMSE. It might be possible to obtain better explanatory power using model terms that are rooted in base expressions for the LRMSE that account for differences in means and other moments, but these expressions are quite complicated, and they induce collinearity with the KL-divergence.[7] While other approaches to the simulation modelling are possible, we believe that the level of explanatory power is sufficient to have confidence in the results. For the pairs of distributions used in the simulation analysis, the results confirm that QS sampling improves the accuracy of importance sampling to estimate a mean quantity from a distribution, and this procedure yields a relative improvement in performance that increases as the sample size for the procedure increases —i.e., the QS importance sampling method works especially well when using larger samples. The results in Figure 5 show a significant increase in accuracy even with relatively small sample sizes, and a very large increase in accuracy for large sample sizes. Our simulation analysis is limited to a stipulated set of test distributions (as it must be), so its generalisability is limited, but it provides further evidence that can be read in conjunction with the previous asymptotic results for QS importance sampling.

## 8. Summary and conclusion

Quantile-stratified (QS) sampling involves sampling from a univariate distribution by first breaking the support up into equiprobable quantile blocks and then sampling one value from each block. This provides an alternative to standard IID sampling, with the stratification ensuring that there is one representative value from each quantile-block in the distribution. The IID and QS samples can be considered as general analogues to SRSWR and SRSWOR and can be expressed in an analogous form where the selected quantile-blocks are sampled with and without replacement. QS sampling can be generalised by allowing "layers" of samples that are combined to yield a layered quantile-stratified (LQS) sample. This generalisation captures IID and QS sampling as extreme cases, with various intermediate cases.

---

[7] We investigated potential model base terms by looking at the LRMSE expression that arises from importance sampling using normal distributions with different means and standard deviations. We also investigated some expressions from other distributional families. In each case these expressions showed collinearity with the KL-divergence in the simulation data, which is not surprising since both the LRMSE and KL-divergence are formed as integrals of functions over the relevant distributions. Although we investigated some expressions of this kind, we made the decision not to use these expressions as base terms for our model, owing to a desire for parsimony and separation of relevant effects. For the model we have used, the difference in means of the distributions enters the analysis only through the KL-divergence term.

The QS sampling method has some advantages over IID sampling for certain problems. The empirical quantiles of the QS samples typically adhere more closely to the true quantiles of the sampling distribution than for an IID sample. Moreover, there is typically less variability in the distances between empirical quantiles in QS samples than in IID samples. This means that QS samples have more stable QQ plots and show a greater adherence to the true quantiles of the sampling distribution. For a uniform sampling distribution there are simple distribution and moment results for the sample values and order statistics in QS samples. The sample values have negative correlation, in contrast to IID samples where they are independent.

Generating samples using QS sampling can be useful in various problems involving estimation of mean quantities through simulation from a stipulated sampling distribution. This includes the general class of problems covered by importance sampling. It is simple to create a variant of standard importance sampling using QS samples from the proposal distribution. This QS importance sampling is typically more accurate than standard importance sampling, owing to the closer adherence of the sample to the true quantiles in the proposal distribution.

**TABLE 1:** Function for quantile-stratified sampling in the **`utilities`** package

| Function | Inputs |
|---|---|
| `qs.sample` | `n, Q, prob.arg = 'p', layers = NULL, ...` |
| **Inputs** | **Inputs** |
| `n` | `The number of sample values to be generated (a non-negative integer)` |
| `Q` | `The quantile function of the sampling distribution (must be a function)` |
| `prob.arg` | `The name of the probability argument in the quantile function Q` |
| `layers` | `Optional vector giving the number of sample values in each layer of the sample` |
| `...` | `Distribution parameters to be passed through to the quantile function Q` |

The QS sample for a univariate distribution is implemented in the **`qs.sample`** function in the **`utilities`** package in **`R`** (O'Neill 2025). Table 1 above shows this function and its inputs. The user must specify the sample size to generate and the quantile function for the sampling distribution.[8] The user can also input distribution parameters that are passed to the quantile function, allowing use with quantile functions for general distributional families (e.g., **`qnorm`**, **`qgamma`**, **`qbeta`**, etc.). The function can also generate LQS samples by adding an input giving the vector of layer sizes for the sample (which must add up to the specified sample size).

In cases where the quantile function for the sampling distribution is unavailable, it is possible to generate a quasi-QS sample from a large enough MCMC sample from the distribution of interest, and this will have similar properties to the true QS sample if the number of values in the MCMC sample is a sufficiently large multiple of the size of the quasi-QS sample.

In this paper we have examined QS sampling for univariate distributions only. It is relatively simple to extend this treatment to the multivariate case, with the requirement that our quantile-blocks would then become regions in higher dimensions and there may be some choices used in the construction of these quantile-blocks. Given a reasonable construction of quantile-blocks in higher dimensions, extension of QS sampling to multivariate distributions will have similarly useful properties and it is left as a topic for future exposition.

**Acknowledgement:** The author would like to thank two anonymous referees for providing helpful suggestions for revision of this paper. The paper has been improved as a result of these suggestions. Any errors remain the responsibility of the author.

[8] By default, the function assumes that the quantile function will use a probability input named **`p`**, which is the standard name in most cases; the user can specify an alternative name for this input using the **`prob.args`** input if it is necessary to accommodate quantile function programmed with a different name for the probability input.

## Declarations

| | |
|---|---|
| **ETHICS** | There was no ethics approval or consent to participate required for this article. |
| **CONSENT FOR PUBLICATION** | This article did not require consent for publication from any person other than the author. |
| **DATA AND MATERIALS** | The following supplementary material was used in this research and is available as supplementary files to this article:<br>- `qs.sample.R`<br>- `Model Data.rds`<br>- `Model Output.rds` |
| **COMPETING INTERESTS** | There are no competing interests to declare. |
| **FUNDING** | The author did not receive support from any organisation for this article. The author has no relevant financial or non-financial interests to disclose. |
| **AUTHOR'S CONTRIBUTION** | The sole author undertook all relevant activities for this research and article. |

## Appendix 1: Proof of Theorems

**PROOF OF THEOREM 1:** The values $S_1^*, \dots, S_n^*$ are obtained from sampling without replacement from the values $1, \dots, n$. As is well-known from sampling theory, we have (taking $i \neq j$):

$$\mathbb{E}(S_i^*) = \frac{n+1}{2},$$
$$\mathbb{V}(S_i^*) = \frac{n^2-1}{12},$$
$$\mathbb{C}(S_i^*, S_j^*) = -\frac{n+1}{12}.$$

Applying the laws of total expectation and variance gives:

$$\begin{aligned}
\mathbb{E}(U_i^*) &= \mathbb{E}(\mathbb{E}(U_i^* | S_i^*)) \\
&= \mathbb{E}\left(\frac{S_i^* - ½}{n}\right) \\
&= \frac{1}{n}(\mathbb{E}(S_i^*) - ½) \\
&= \frac{1}{n}\left(\frac{n+1}{2} - ½\right) \\
&= \frac{1}{n} \cdot \frac{n}{2} \\
&= \frac{1}{2},
\end{aligned}$$

$$\begin{aligned}
\mathbb{V}(U_i^*) &= \mathbb{E}(\mathbb{V}(U_i^* | S_i^*)) + \mathbb{V}(\mathbb{E}(U_i^* | S_i^*)) \\
&= \mathbb{E}\left(\frac{1}{12n^2}\right) + \mathbb{V}\left(\frac{S_i^* - ½}{n}\right) \\
&= \frac{1}{12n^2} + \frac{n^2-1}{12n^2} \\
&= \frac{1}{12}.
\end{aligned}$$

For all $i \neq j$, applying the law of total covariance gives:

$$\begin{aligned}
\mathbb{C}(U_i^*, U_j^*) &= \mathbb{E}(\mathbb{C}(U_i^*, U_j^* | S_i^*, S_j^*)) + \mathbb{C}(\mathbb{E}(U_i^* | S_i^*, S_j^*), \mathbb{E}(U_j^* | S_i^*, S_j^*)) \\
&= \mathbb{E}(0) + \mathbb{C}\left(\frac{S_i^* - ½}{n}, \frac{S_j^* - ½}{n}\right) \\
&= 0 + \frac{1}{n^2}\mathbb{C}(S_i^*, S_j^*)
\end{aligned}$$

$$= -\frac{1}{n^2}\frac{n+1}{12}$$
$$= -\frac{n+1}{12n^2}.$$

The correlation shown in the theorem easily follows, which completes the proof. ■

**PROOF OF THEOREM 2:** For each value $U_i^{**}$ in the layered sample, let $W_i$ denote the subsample from which it was generated (i.e., the first element of the generated value $S_i^{**}$). Since each of the subsamples is a QS sample, they have the fixed mean and variance shown in Theorem 1. Applying the laws of total expectation and variance therefore gives:

$$\mathbb{E}(U_i^*) = \mathbb{E}(\mathbb{E}(U_i^*|W_i))$$
$$= \mathbb{E}\left(\frac{1}{2}\right)$$
$$= \frac{1}{2},$$

$$\mathbb{V}(U_i^{**}) = \mathbb{E}(\mathbb{V}(U_i^{**}|W_i)) + \mathbb{V}(\mathbb{E}(U_i^{**}|W_i))$$
$$= \mathbb{E}\left(\frac{1}{12}\right) + \mathbb{V}(\mu)$$
$$= \frac{1}{12}.$$

Moreover, applying the covariance result in Theorem 1 we get:

$$\mathbb{C}(U_i^{**}, U_j^{**}|W_i = W_j = k) = -\frac{n_k+1}{12n_k^2} \qquad \text{for } k = 1, \dots, K.$$

Applying this result and using the law of total covariance we then have:

$$\mathbb{C}(U_i^{**}, U_j^{**}) = \mathbb{E}(\mathbb{C}(U_i^{**}, U_j^{**}|W_i, W_j)) + \mathbb{C}(\mathbb{E}(U_i^{**}|W_i), \mathbb{E}(U_j^{**}|W_j))$$
$$= \sum_{k=1}^{K} \mathbb{C}(U_i^{**}, U_j^{**}|W_i = W_j = k) \cdot \mathbb{P}(W_i = W_j = k) + \mathbb{Cov}(\mu, \mu)$$
$$= -\sum_{k=1}^{K} \frac{n_k+1}{12n_k^2} \cdot \frac{n_k}{n} \cdot \frac{n_k-1}{n-1}$$
$$= -\frac{1}{12n(n-1)} \sum_{k=1}^{K} \frac{(n_k+1)(n_k-1)}{n_k}$$
$$= -\frac{1}{12n(n-1)} \sum_{k=1}^{K} \frac{n_k^2-1}{n_k}$$

$$= -\frac{1}{12n(n-1)} \sum_{k=1}^{K} \left[ n_k - \frac{1}{n_k} \right]$$

$$= -\frac{n - \sum_{k=1}^{K} 1/n_k}{12n(n-1)}.$$

The correlation shown in the theorem easily follows, which completes the proof. ■

**PROOF OF THEOREM 3:** The order statistics for the two methods have marginal distributions:

$$U_{(k)} \sim \text{Beta}(k, n-k+1),$$

$$U_{(k)}^* \sim \text{U}\left(\frac{k-1}{n}, \frac{k}{n}\right).$$

Using the moment equations from these distributions, the mean and variance results are:

$$\mathbb{E}(U_{(k)}) = \frac{k}{n+1} = p_k,$$

$$\mathbb{V}(U_{(k)}) = \frac{1}{n+2} \cdot \frac{k}{n+1} \cdot \frac{n-k+1}{n+1} = \frac{p_k(1-p_k)}{n+2},$$

$$\mathbb{E}(U_{(k)}^*) = \frac{2k-1}{2n} = \frac{k - ½}{n} = p_k^*,$$

$$\mathbb{V}(U_{(k)}^*) = \frac{1}{12} \cdot \frac{1}{n^2} = \frac{1}{12n^2}.$$

Since $U_{(k)}$ is an unbiased estimator of $p_k$ and $U_{(k)}^*$ is an unbiased estimator of $p_k^*$ we have the following simple MSE results:

$$\begin{aligned}
\text{MSE}_{n,k}(p_k) &= \mathbb{E}((U_{(k)} - p_k)^2) \\
&= \mathbb{V}(U_{(k)}) + \text{Bias}(U_{(k)}|p_k)^2 \\
&= \frac{p_k(1-p_k)}{n+2} + 0^2 \\
&= \frac{p_k(1-p_k)}{n+2}, \\
\text{MSE}_{n,k}^*(p_k^*) &= \mathbb{E}((U_{(k)}^* - p_k^*)^2) \\
&= \mathbb{V}(U_{(k)}^*) + \text{Bias}(U_{(k)}^*|p_k^*)^2 \\
&= \frac{1}{12n^2} + 0^2 \\
&= \frac{1}{12n^2}.
\end{aligned}$$

The remaining two MSE results are a bit more complicated because they concern estimators that are biased. As a preliminary step, with a bit of algebra, it can be shown that:

$$p_k^* - p_k = \frac{p_k - ½}{n} = \frac{p_k^* - ½}{n+1}.$$

We therefore have:

$$\begin{aligned}
\text{MSE}_{n,k}^*(p_k) &= \mathbb{E}((U_{(k)}^* - p_k)^2) \\
&= \mathbb{V}(U_{(k)}^*) + \text{Bias}(U_{(k)}^*|p_k)^2 \\
&= \frac{1}{12n^2} + (p_k^* - p_k)^2 \\
&= \frac{1}{3n^2} - \frac{1}{4n^2} + \frac{(p_k - ½)^2}{n^2} \\
&= \frac{1}{3n^2} - \frac{1 - 4(p_k - ½)^2}{4n^2} \\
&= \frac{1}{3n^2} - \frac{1 - (4p_k^2 - 4p_k + 1)}{4n^2} \\
&= \frac{1}{3n^2} - \frac{(p_k^2 - p_k)}{n^2} \\
&= \frac{1}{3n^2} - \frac{p_k(1 - p_k)}{n^2},
\end{aligned}$$

$$\begin{aligned}
\text{MSE}_{n,k}(p_k^*) &= \mathbb{E}((U_{(k)} - p_k^*)^2) \\
&= \mathbb{V}(U_{(k)}) + \text{Bias}(U_{(k)}|p_k^*)^2 \\
&= \frac{p_k(1 - p_k)}{n+2} + (p_k^* - p_k)^2 \\
&= \frac{k(n-k+1)}{(n+1)^2(n+2)} + \left(\frac{p_k^* - ½}{n+1}\right)^2 \\
&= \frac{k(n-k+1)}{(n+1)^2(n+2)} + \frac{(p_k^* - ½)^2}{(n+1)^2} \\
&= \frac{k(n-k+1) + (n+2)(p_k^* - ½)^2}{(n+1)^2(n+2)} \\
&= \frac{k(n+1) - k^2 + (n+2)(p_k^* - ½)^2}{(n+1)^2(n+2)} \\
&= \frac{(k - ½)n - (k - ½)^2 + ½(n + ½) + (n+2)(p_k^* - ½)^2}{(n+1)^2(n+2)} \\
&= \frac{n^2 p_k^* - n^2 p_k^{*2} + ½(n + ½) + (n+2)(p_k^{*2} - p_k^* + ¼)}{(n+1)^2(n+2)}
\end{aligned}$$

$$= \frac{-(n^2 - n - 2)p_k^{*2} + (n^2 - n - 2)p_k^* + ¾(n+1)}{(n+1)^2(n+2)}$$

$$= \frac{-(n-2)p_k^{*2} + (n-2)p_k^* + ¾}{(n+1)(n+2)}$$

$$= \frac{(n-2)p_k^*(1-p_k^*) + ¾}{(n+1)(n+2)}.$$

This establishes each of the MSE results and completes the proof. ■

**PROOF OF THEOREM 4:** The beta distribution for $D_{k,\ell}$ is a well-known result for order statistics (see e.g., Reiss 1989, pp. 21-22) and so are its moments, so we omit the derivation here. The triangular distribution for $D_{k,\ell}^*$ (see Kotz and Van Dorpe 2004) is obtained as:

$$p(D_{k,\ell}^* = d) = p(U_{(k+\ell)}^* - U_{(k)}^* = d)$$

$$= \int_0^1 p(U_{(k+\ell)}^* = u + d) \cdot p(U_{(k)}^* = u) du$$

$$= n^2 \int_0^1 \mathbb{I}\left(\frac{k+\ell-1}{n} < u + d \le \frac{k+\ell}{n}\right) \cdot \mathbb{I}\left(\frac{k-1}{n} < u \le \frac{k}{n}\right) du$$

$$= n^2 \int_0^1 \mathbb{I}\left(\frac{k+\ell-1}{n} - \min\left(\frac{\ell}{n}, d\right) < u \le \frac{k}{n} - d + \min\left(\frac{\ell}{n}, d\right)\right) du$$

$$= \begin{cases} n^2 \int_0^1 \mathbb{I}\left(\frac{k+\ell-1}{n} - d < u \le \frac{k}{n}\right) du & \frac{\ell-1}{n} \le d \le \frac{\ell}{n} \\ n^2 \int_0^1 \mathbb{I}\left(\frac{k-1}{n} < u \le \frac{k+\ell}{n} - d\right) du & \frac{\ell}{n} \le d \le \frac{\ell+1}{n} \\ 0 & \text{otherwise} \end{cases}$$

$$= \begin{cases} n^2\left(d - \frac{\ell-1}{n}\right) & \frac{\ell-1}{n} \le d \le \frac{\ell}{n} \\ n^2\left(\frac{\ell+1}{n} - d\right) & \frac{\ell}{n} \le d \le \frac{\ell+1}{n} \\ 0 & \text{otherwise} \end{cases}$$

$$= \text{Triangular}\left(d \middle| \frac{\ell-1}{n}, \frac{\ell}{n}, \frac{\ell+1}{n}\right).$$

Using standard formulae for the moments of this distribution (Kotz and Van Dorpe 2004, pp. 8-11) we then have:

$$\begin{aligned}\mathbb{E}(D^*_{k,\ell}) &= \frac{1}{3}\left[\left(\frac{\ell-1}{n}\right)+\left(\frac{\ell}{n}\right)+\left(\frac{\ell+1}{n}\right)\right] \\ &= \frac{1}{3n}[(\ell-1)+(\ell)+(\ell+1)] \\ &= \frac{3\ell}{3n}=\frac{\ell}{n},\end{aligned}$$

$$\begin{aligned}\mathbb{V}(D^*_{k,\ell}) &= \frac{1}{18}\begin{bmatrix} \left(\frac{\ell-1}{n}\right)^2+\left(\frac{\ell}{n}\right)^2+\left(\frac{\ell+1}{n}\right)^2 \\ -\left(\frac{\ell-1}{n}\right)\left(\frac{\ell}{n}\right)-\left(\frac{\ell-1}{n}\right)\left(\frac{\ell+1}{n}\right)-\left(\frac{\ell}{n}\right)\left(\frac{\ell+1}{n}\right)\end{bmatrix} \\ &= \frac{1}{18n^2}\begin{bmatrix}(\ell^2-2\ell+1)+(\ell^2)+(\ell^2+2\ell+1) \\ -(\ell^2-\ell)-(\ell^2-1)-(\ell^2+\ell)\end{bmatrix} \\ &= \frac{3}{18n^2}=\frac{1}{6n^2}.\end{aligned}$$

This establishes the moments in the theorem for the triangular distribution, and we defer to the cited literature for the remaining parts of the theorem. ■

**PROOF OF THEOREM 5:** Under the initial assumption of the theorem (that the order statistics in the MCMC chain follow the standard distribution for the order statistics from an IID sample) we have:

$$F(M_{(i)}) \sim \text{Beta}(i, nr-i+1) \qquad i=1,\dots,nr.$$

Since $U^*_{(k)} = F(X^*_{(k)}) = F(M_{((k-1)r+T)})$ with $T \sim \text{U}\{1,\dots,r\}$, the distribution of this value is a uniform mixture of beta distributions, so it has density given by:

$$p(U^*_{(s)}=u) = \frac{1}{r}\sum_{t=1}^{r}\text{Beta}(u|(s-1)r+t,(n-s+1)r-t+1).$$

Applying the law of total expectation we obtain:

$$\begin{aligned}\mathbb{E}(U^*_{(k)}) &= \mathbb{E}(\mathbb{E}(U^*_{(k)}|T)) \\ &= \frac{1}{r}\sum_{t=1}^{r}\mathbb{E}(U^*_{(k)}|T=t) \\ &= \frac{1}{r}\sum_{t=1}^{r}\frac{(k-1)r+t}{nr+1} \\ &= \frac{1}{nr+1}\left[(k-1)r+\frac{1}{r}\sum_{t=1}^{r}t\right]\end{aligned}$$

$$= \frac{1}{nr+1}\left[(k-1)r + \frac{r+1}{2}\right]$$

$$= \frac{(k-½)r+½}{nr+1},$$

$$\mathbb{E}(U_{(k)}^{*2}) = \mathbb{E}(\mathbb{E}(U_{(k)}^{*2}|T))$$

$$= \frac{1}{r}\sum_{t=1}^{r}\mathbb{E}(U_{(k)}^{*2}|T=t)$$

$$= \frac{1}{r}\sum_{t=1}^{r}\frac{(k-1)r+t}{nr+1}\cdot\frac{(k-1)r+t+1}{nr+2}$$

$$= \frac{1}{(nr+1)(nr+2)}\frac{1}{r}\sum_{t=1}^{r}[(k-1)r+t][(k-1)r+t+1]$$

$$= \frac{1}{(nr+1)(nr+2)}\frac{1}{r}\sum_{t=1}^{r}[(k-1)^2r^2+(k-1)r+2(k-1)rt+t+t^2]$$

$$= \frac{1}{(nr+1)(nr+2)}\begin{bmatrix}(k-1)^2r^2+(k-1)r \\ +(k-1)r(r+1)+\frac{1}{2}(r+1) \\ +\frac{(r+1)(2r+1)}{6}\end{bmatrix}$$

$$= \frac{1}{(nr+1)(nr+2)}\begin{bmatrix}(k-1)^2r^2+(k-1)r^2+\frac{1}{3}r^2 \\ +(k-1)r+(k-1)r+\frac{1}{2}r+\frac{1}{2}r \\ +\frac{1}{2}+\frac{1}{6}\end{bmatrix}$$

$$= \frac{1}{(nr+1)(nr+2)}\left[\left(k(k-1)+\frac{1}{3}\right)r^2+2\left(k-\frac{1}{2}\right)r+\frac{2}{3}\right]$$

$$= \frac{(k(k-1)+⅓)r^2+2(k-½)r+⅔}{(nr+1)(nr+2)}.$$

$$\mathbb{V}(U_{(k)}^{*}) = \mathbb{E}(U_{(s)}^{*2}) - \mathbb{E}(U_{(s)}^{*})^2$$

$$= \frac{(k(k-1)+⅓)r^2+2(k-½)r+⅔}{(nr+1)(nr+2)} - \frac{[(k-½)r+½]^2}{(nr+1)^2}$$

$$= \frac{(k(k-1)+⅓)r^2+2(k-½)r+⅔}{(nr+1)(nr+2)} - \frac{(k-½)^2r^2+(k-½)r+¼}{(nr+1)^2}$$

$$= \frac{(nr+1)[(k(k-1)+⅓)r^2+2(k-½)r+⅔] \\ -(nr+2)[(k-½)^2r^2+(k-½)r+¼]}{(nr+1)^2(nr+2)}$$

$$= \frac{\begin{bmatrix} n(k(k-1)+⅓)r^3 + 2n(k-½)r^2 + ⅔nr \\ +(k(k-1)+⅓)r^2 + 2(k-½)r + ⅔ \\ -n(k-½)^2r^3 - n(k-½)r^2 - ¼nr \\ -2(k-½)^2r^2 - 2(k-½)r - ½ \end{bmatrix}}{(nr+1)^2(nr+2)}$$

$$= \frac{\begin{bmatrix} n(k(k-1)+⅓)r^3 - n(k-½)^2r^3 \\ +2n(k-½)r^2 + (k(k-1)+⅓)r^2 - n(k-½)r^2 - 2(k-½)^2r^2 \\ +⅔nr + 2(k-½)r - ¼nr - 2(k-½)r \\ +⅔ - ½ \end{bmatrix}}{(nr+1)^2(nr+2)}$$

$$= \frac{\frac{n}{12}r^3 + \left[n(k-½) - k^2 + k - \frac{1}{6}\right]r^2 + \frac{5}{12}nr + \frac{1}{6}}{(nr+1)^2(nr+2)}$$

$$= \frac{nr^3 + [12n(k-½) - 12k(k-1) - 2]r^2 + 5nr + 2}{12(nr+1)^2(nr+2)}.$$

We then have mean-squared error given by:

$$\begin{aligned} \text{MSE}^*_{n,k}(p^*_k) &= \mathbb{E}((U^*_{(k)} - p^*_k)^2) \\ &= \mathbb{E}(U^{*2}_{(k)} - 2p_kU^*_{(k)} + p^2_k) \\ &= \mathbb{E}(U^{*2}_{(k)}) - 2p_k\,\mathbb{E}(U^*_{(k)}) + p^2_k \end{aligned}$$

$$= \frac{(k(k-1)+⅓)r^2 + 2(k-½)r + ⅔}{(nr+1)(nr+2)} - 2\frac{k-½}{n}\frac{(k-½)r+½}{nr+1} + \left(\frac{k-½}{n}\right)^2$$

$$= \frac{(k(k-1)+⅓)r^2 + 2(k-½)r + ⅔}{(nr+1)(nr+2)} - \frac{2(k-½)[(k-½)r+½]}{n(nr+1)} + \frac{(k-½)^2}{n^2}$$

$$= \frac{\begin{bmatrix} n^2(k(k-1)+⅓)r^2 + 2n^2(k-½)r + ⅔n^2 \\ -2n(nr+2)(k-½)[(k-½)r+½] \\ +(nr+1)(nr+2)(k-½)^2 \end{bmatrix}}{n^2(nr+1)(nr+2)}$$

$$= \frac{\begin{bmatrix} n^2(k(k-1)+⅓)r^2 + 2n^2(k-½)r + ⅔n^2 \\ -2n^2(k-½)^2r^2 - 4n(k-½)^2r - n^2(k-½)r - 2(k-½)n \\ +(k-½)^2(n^2r^2 + 3nr + 2) \end{bmatrix}}{n^2(nr+1)(nr+2)}$$

$$= \frac{\begin{bmatrix} n^2(k(k-1)+⅓)r^2 - 2n^2(k-½)^2r^2 + n^2(k-½)^2r^2 \\ +2n^2(k-½)r - 4n(k-½)^2r - n^2(k-½)r + 3n(k-½)^2r \\ +⅔n^2 - 2(k-½)n + 2(k-½)^2 \end{bmatrix}}{n^2(nr+1)(nr+2)}$$

$$= \frac{n^2r^2 + 12n(k-½)(n-k+½)r + 8n^2 - 24(k-½)n + 24(k-½)^2}{12n^2(nr+1)(nr+2)}.$$

This establishes the results in the theorem. ■

## Appendix 2: Details of simulation analysis

This appendix contains details of the simulation analysis examining the accuracy of estimates in importance sampling using IID samples and QS samples. In Table 2 below we show the twenty distributions that were used for the simulation (with their mean and standard deviation).

**TABLE 2: Distributions used in simulation analysis**

| Distribution | Mean | SD |
|---|---|---|
| Gamma (shape = 2.0, scale = 1.0) | 2.0000 | 1.4142 |
| Gamma (shape = 2.4, scale = 1.2) | 2.8800 | 1.8590 |
| Gamma (shape = 2.3, scale = 0.8) | 1.8400 | 1.2133 |
| Gamma (shape = 2.1, scale = 0.9) | 1.8900 | 1.3042 |
| Gamma (shape = 1.8, scale = 1.1) | 1.9800 | 1.4758 |
| Scaled F ($df_1$ = 6.0, $df_2$ = 8.0, scale = 1.2) | 1.6000 | 1.6000 |
| Scaled F ($df_1$ = 6.0, $df_2$ = 9.0, scale = 1.3) | 1.6714 | 1.5560 |
| Scaled F ($df_1$ = 5.0, $df_2$ = 8.0, scale = 1.3) | 1.7333 | 1.8179 |
| Scaled F ($df_1$ = 6.0, $df_2$ = 7.0, scale = 1.4) | 1.9600 | 2.1669 |
| Scaled F ($df_1$ = 7.0, $df_2$ = 9.0, scale = 1.3) | 1.6714 | 1.4950 |
| Scaled lognormal (meanlog = 0.5, sdlog = 0.7, scale = 0.8) | 1.6852 | 1.3400 |
| Scaled lognormal (meanlog = 0.6, sdlog = 0.8, scale = 0.7) | 1.7565 | 1.6631 |
| Scaled lognormal (meanlog = 0.4, sdlog = 0.7, scale = 0.9) | 1.7154 | 1.3641 |
| Scaled lognormal (meanlog = 0.5, sdlog = 0.8, scale = 0.8) | 1.8164 | 1.7198 |
| Scaled lognormal (meanlog = 0.5, sdlog = 0.6, scale = 1.0) | 1.9739 | 1.2994 |
| Weibull (shape = 1.3, scale = 2.0) | 1.8472 | 1.4329 |
| Weibull (shape = 1.5, scale = 1.9) | 1.7152 | 1.1646 |
| Weibull (shape = 1.4, scale = 1.8) | 1.6406 | 1.1874 |
| Weibull (shape = 1.3, scale = 2.2) | 2.0319 | 1.5762 |
| Weibull (shape = 1.4, scale = 2.1) | 1.9140 | 1.3853 |

We used a Gaussian linear regression model. The regression model has the form:

$$\text{LRMSE}(i, j, n, \text{QS}) = \log\left(\frac{\sigma_i}{\sqrt{n}}\right) + \beta_0 + \beta_1 \log\left(\frac{\sigma_j}{\sigma_i}\right) + \beta_3 \log(KL_{i,j})$$

$$+\beta_4 \log(n) + \beta_5 \text{QS} \log(n) + \text{Error}.$$

The output of the regression analysis is shown in the regression tables below. This includes standard output tables for the model, plus a table of the estimated multiplicative effects.

| Summary Statistics | |
|---|---|
| Multiple R | 0.8033 |
| R-square | 0.6453 |
| Adj. R-square | 0.6450 |
| Standard Error | 1.0496 |
| Observations | 4000 |

**ANOVA Table**

| Component | SS | df | MS | F | p |
|---|---|---|---|---|---|
| Standard-deviation-ratio | 4779.44 | 1 | 4779.44 | 4352.36 | 0.0000 |
| KL divergence | 271.35 | 1 | 271.35 | 246.08 | 0.0000 |
| Log-sample-size | 145.42 | 1 | 145.42 | 131.87 | 0.0000 |
| QS sampling:Log-sample-size | 2798.67 | 1 | 2798.67 | 2537.97 | 0.0000 |
| Residual | 4405.38 | 3995 | 1.10 | | |
| Total | 12420.26 | 3999 | 3.11 | | |

**Coefficient Estimates Table**

| Component | Coef | SE | t | p | 95% Conf Interval | |
|---|---|---|---|---|---|---|
| (Intercept) | 3.33680 | 0.05190 | 64.288 | 0.0000 | 3.23506 | 3.43852 |
| Standard-deviation-ratio | 5.07570 | 0.07541 | 67.306 | 0.0000 | 4.92789 | 5.22350 |
| KL divergence | 2.72607 | 0.17378 | 15.687 | 0.0000 | 2.38547 | 3.06668 |
| Log-sample-size | 0.05485 | 0.01075 | 5.102 | 0.0000 | 0.03378 | 0.07593 |
| QS sampling:Log-sample-size | -0.34389 | 0.00683 | -50.378 | 0.0000 | -0.35727 | -0.33051 |

**Estimated Multiplicative Effects (by sample size and sampling type)[9]**

| m | 10 | 30 | 100 | 300 | 1000 |
|---|---|---|---|---|---|
| IID sampling | 1.13463 | 1.20511 | 1.28739 | 1.36735 | 1.46070 |
| QS sampling | 0.51400 | 0.37416 | 0.26420 | 0.19232 | 0.13580 |
| Ratio (QS/IID) | 0.45301 | 0.31048 | 0.20522 | 0.14065 | 0.09297 |

The above model was used to generate the residuals and studentised residuals for each of the data points in the simulation dataset. We examined diagnostic plots of the model residuals to determine whether there was deviation from the assumed model form that would require model variation. Our diagnostic analysis showed some evidence of a quadratic trend in residuals.[10] We also saw evidence of positive skew and a heavy right-tail in the residuals. These diagnostic results are exhibited in the residual plot and QQ plot for the studentised residuals shown below in Figure 6. We also examined a scale-location plot and a leverage plot for the studentised residuals, but we have omitted these for brevity.

[9] The multiplicative effects in this table are for the estimated RMSE of the importance sampling estimators. They are calculated from the estimated coefficients in the model. The estimates are given by $\hat{\phi}_{\text{IID},n} \equiv \exp(\hat{\beta}_4 \log(n))$ and $\hat{\phi}_{\text{QS},n} \equiv \exp((\hat{\beta}_4 + \hat{\beta}_5) \log(n))$ for the two sampling types. The resulting ratios of these multiplicative effects follow directly.
[10] This may be due to some omitted variable that would aid in explanation of the LRMSE.

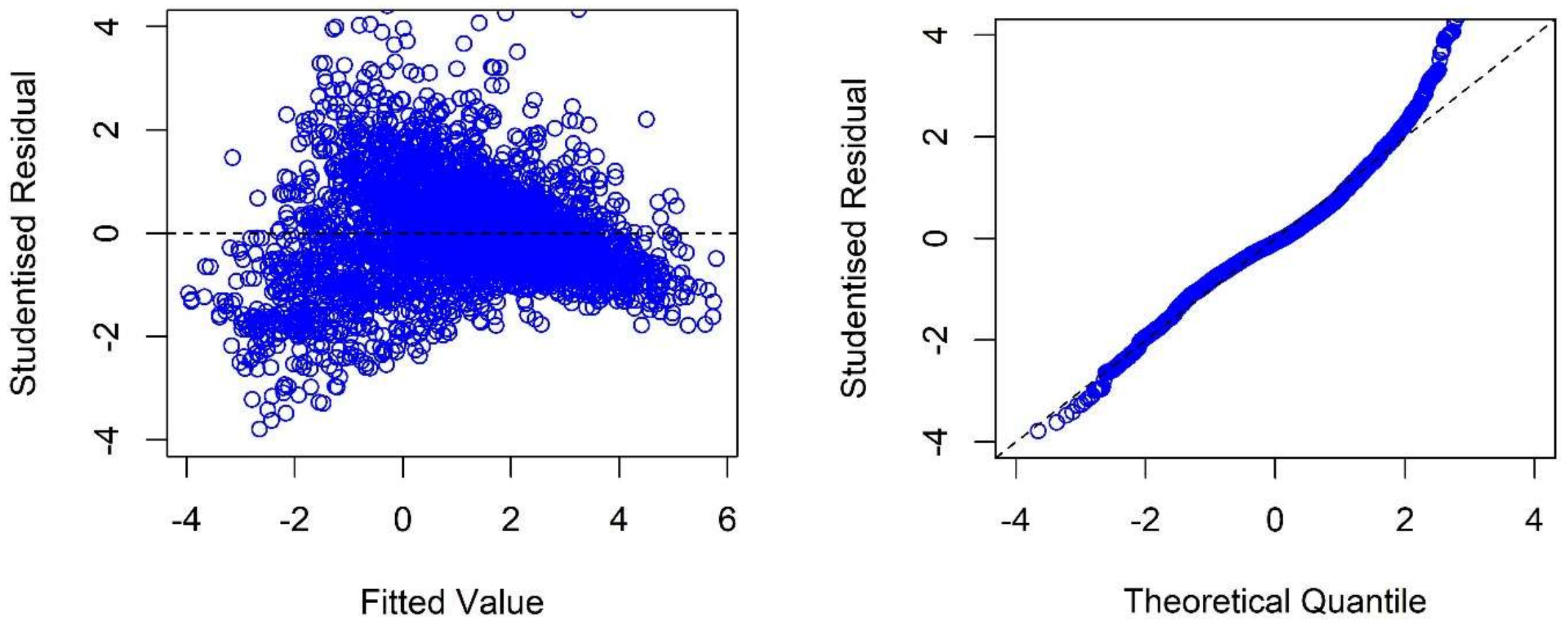

**FIGURE 6:** Residual plot and QQ plot of studentised residuals

Notwithstanding some evident departures from model assumptions, the explanatory power of the model was reasonable and the results for the explanatory variables were in line with a priori expectations. Estimates of the model coefficients are robust to deviations from error-normality and heteroskedasticity, but they are affected by nonlinearity in the model. The estimates of the relative effects of IID sampling versus QS sampling on the LRMSE may be affected by this and might be more accurately estimated in a more detailed model.

All simulation analysis in this paper was conducted in **`R`**. The model data for this analysis is available in the file **`Model Data.rds`** and the corresponding model output is available in the file **`Model Output.rds`** (available in supplementary files).